\documentclass[12pt]{article}
\usepackage[dvips]{graphicx}
\def\thefootnote{\fnsymbol{footnote}}
\def\be{\begin{equation}}
\def\ee{\end{equation}}
\def\ba{\begin{eqnarray}}
\def\ea{\end{eqnarray}}

\begin{document}
\begin{titlepage}
\thispagestyle{empty}

\vskip0.2cm
\begin{flushright}
MS--TPI--98--17
\end{flushright}
\vskip0.8cm             

\begin{center}
{\Large {\bf A Chaotic Cousin}}
\end{center}

\begin{center}
{\Large {\bf Of Conway's Recursive Sequence}}
\end{center}
\vskip0.5cm
\begin{center}
{\large Klaus Pinn}\\
\vskip5mm
{Institut f\"ur Theoretische Physik I } \\
{Universit\"at M\"unster }\\ {Wilhelm--Klemm--Str.~9 }\\
{D--48149 M\"unster, Germany \\[5mm]
 e--mail: pinn@uni--muenster.de
 }
\end{center}
\vskip0.5cm
\begin{abstract}
\par\noindent
I study the recurrence $D(n)= D(D(n-1))+D(n-1-D(n-2))$, $D(1)=D(2)=1$.
Its definition has some similarity to that of Conway's sequence
defined through $a(n)= a(a(n-1))+a(n-a(n-1))$, $a(1)=a(2)=1$.
However, in contradistinction to the completely regular and
predictable behaviour of $a(n)$, the $D$-numbers exhibit chaotic
patterns. 
In its statistical properties, the $D$-sequence shows striking
similarities with Hofstadter's $Q(n)$-sequence, defined through $Q(n)=
Q(n-Q(n-1))+Q(n-Q(n-2))$, $Q(1)=Q(2)=1$.  Compared to the Hofstadter
sequence, the $D$-recurrence shows higher structural order.  It is
organized in well-defined ``generations'', separated by smooth and
predictable regions. 
The article is complemented by a study of two further recurrence
relations with definitions similar to those of the $Q$-numbers. 
There is some evidence that the different sequences studied 
share a universality class. Could it be that there are some real 
life processes modelled by these recurrences? \\[4mm]
I OFFER A CASH PRIZE OF \$100 TO THE FIRST PROVIDING
A PROOF OF SOME CONJECTURES ABOUT $D(n)$ FORMULATED IN THIS ARTICLE. 

\end{abstract}
\end{titlepage}

\setcounter{footnote}{0} \def\thefootnote{\arabic{footnote}}

\section{Introduction}

\begin{figure}
\begin{center}
\includegraphics[height=7cm]{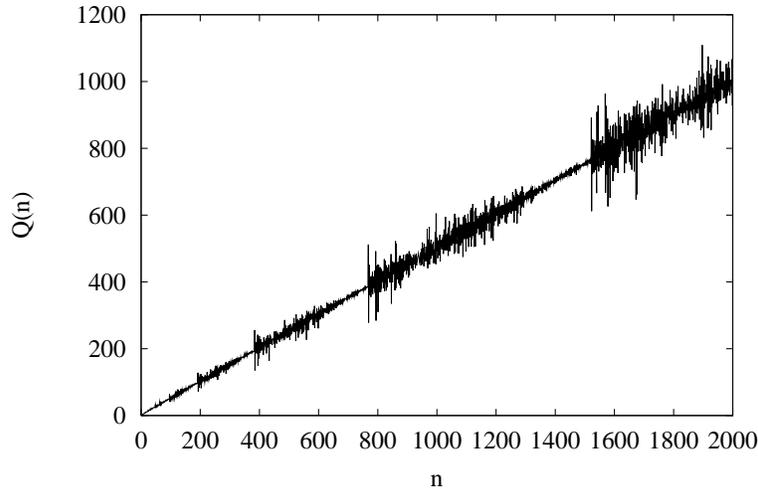}
\parbox[t]{.85\textwidth}
 {
 \caption[appearQ]
 {\label{appearQ}
\small
 Graph of $Q(n)$.
}}
\end{center}
\end{figure}

The recursion relation  
\be\label{QQQ}
\begin{array}{ll}
 & Q(n) = Q(n-Q(n-1)) + Q(n-Q(n-2)) \quad \mbox{for} ~~ n > 2 \, , \\[3mm]
 & Q(1) = Q(2) = 1 \, , 
\end{array}
\ee 
introduced by D. R. Hofstadter in his book {\sc G\"odel, Escher, Bach:
an Eternal Golden Braid} \cite{GEB}, is a challenge \cite{Guy}.  Its
apparently chaotic behaviour (see figure \ref{appearQ}) is far from
being understood. There appear to be no rigorous results about the
behaviour of $Q(n)$.

In ref.~\cite{mypaper} I reported a number of mainly statistical 
observations on the $Q$-numbers. The main conclusions were: 

\begin{itemize}
\item The sequence shows some signals of order. It is 
organized in ``generations'', 
making up for a Fibonacci-type structure on a logarithmic scale.
\item The variance of fluctuations around $n/2$ 
goes like $n^{\alpha}$, with $\alpha= 0.88(1)$.
\item $R(n)= (Q(n)-n/2)/n^\alpha$ has a well-defined, strongly
non-Gaussian probability density $p^*$. 
\item There is scaling: 
$x_m = R(n)-R(n-m)$ is distributed according to 
$\lambda_m p^*(x_m/\lambda_m)$. The rescaling 
factor $\lambda_m$ converges to $\sqrt{2}$ for large $m$, 
exponentially fast with a decay length $\xi = 3$.
\end{itemize} 

It is an interesting question whether similar observations 
can be made on other integer recurrences. 
In this paper, I introduce and study the recurrence 
\be\label{DDD}
\begin{array}{ll}
 & D(n) = D(D(n-1)) + D(n-1-D(n-2)) \quad \mbox{for} ~~ n > 2 \, ,  \\[3mm]
 & D(1) = D(2) = 1 \, .
\end{array}
\ee  
Its definition is not too different from 
that of 
Conway's sequence $a(n)$, defined through
\be\label{aaa}
\begin{array}{ll}
 & a(n) = a(a(n-1)) + a(n-a(n-1)) \quad \mbox{for} ~~ n > 2 \, , \\[3mm]
 & a(1) = a(2) = 1 \, .
\end{array}
\ee 
The $a(n)$-sequence has been investigated by Hofstadter, Conway and
others at various times since about 1975 \cite{Conolly,HofCon}.
Conway discovered many of its properties. A cash prize that he offered
for information about its asymptotic growth was won by Mallows
\cite{Mallows}. See also \cite{Kubo} for a detailed study of $a(n)$.

Conway's sequence has a lot of fascinating properties. However, it
behaves in a regular and completely predictable way.  In contrast,
$D(n)$ develops chaotic and irregular patterns, separated by smooth
and predictable regions. The latter property underlines its close
relation to the $a(n)$ function.

In section 2, I report on some non-statistical observations of
properties of the $D$-sequence.  Section 3 is about some of its
statistical properties. The section describes investigations of the
step size distribution, scaling properties, and frequency counting.
Section 4 complements the study of the $D$-sequence by an
investigation of two further chaotic recurrences that might be called
chaotic cousins of the Hofstadter sequence.  All sequences studied
share various statistical properties. This suggests that they belong
to a common universality class. Because of its clear structure the
$D$-sequence seems to be a natural candidate for studies of this
class.

\section{Non-Statistical Observations} 

\begin{figure}
\begin{center}
\includegraphics[height=8cm]{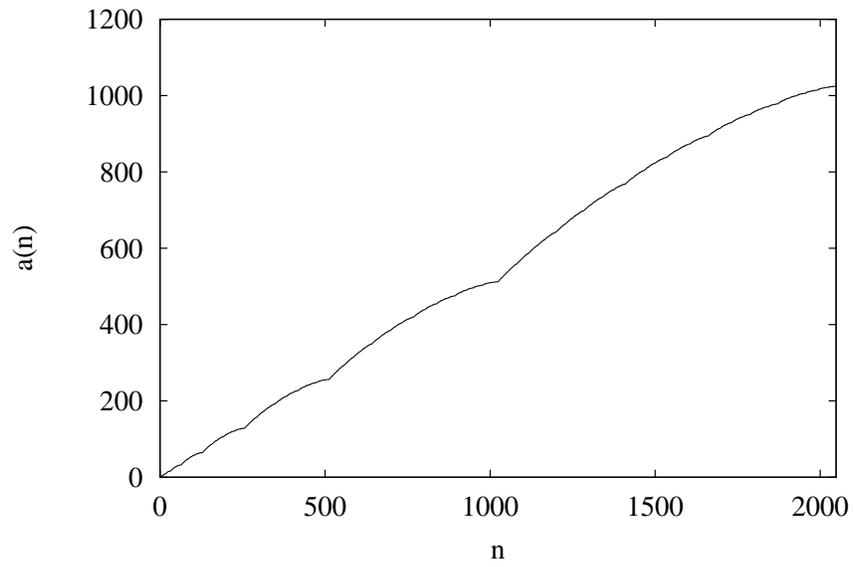}

\includegraphics[height=8cm]{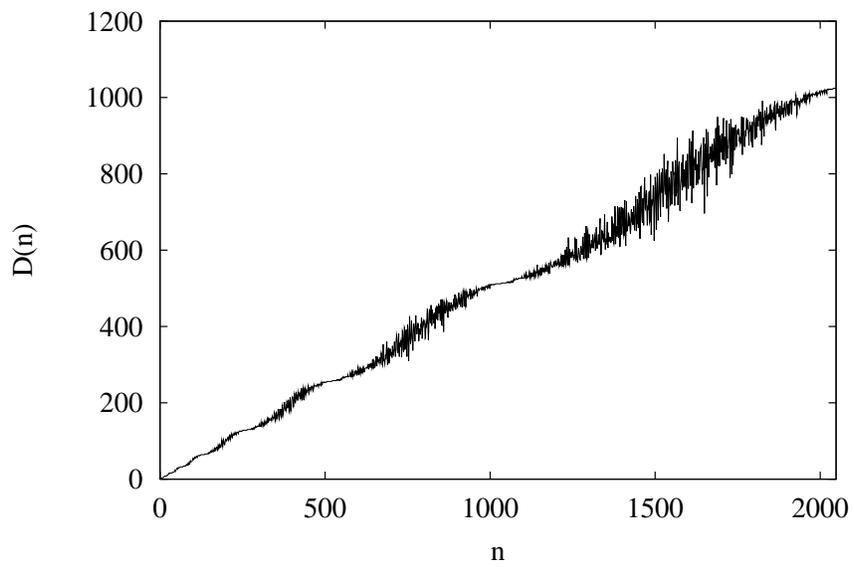}

\parbox[t]{.85\textwidth}
 {
 \caption[appear1]
 {\label{appear1}
\small
 Graphs of $a(n)$ and $D(n)$.
}}
\end{center}
\end{figure}

\begin{figure}
\begin{center}
\includegraphics[height=8cm]{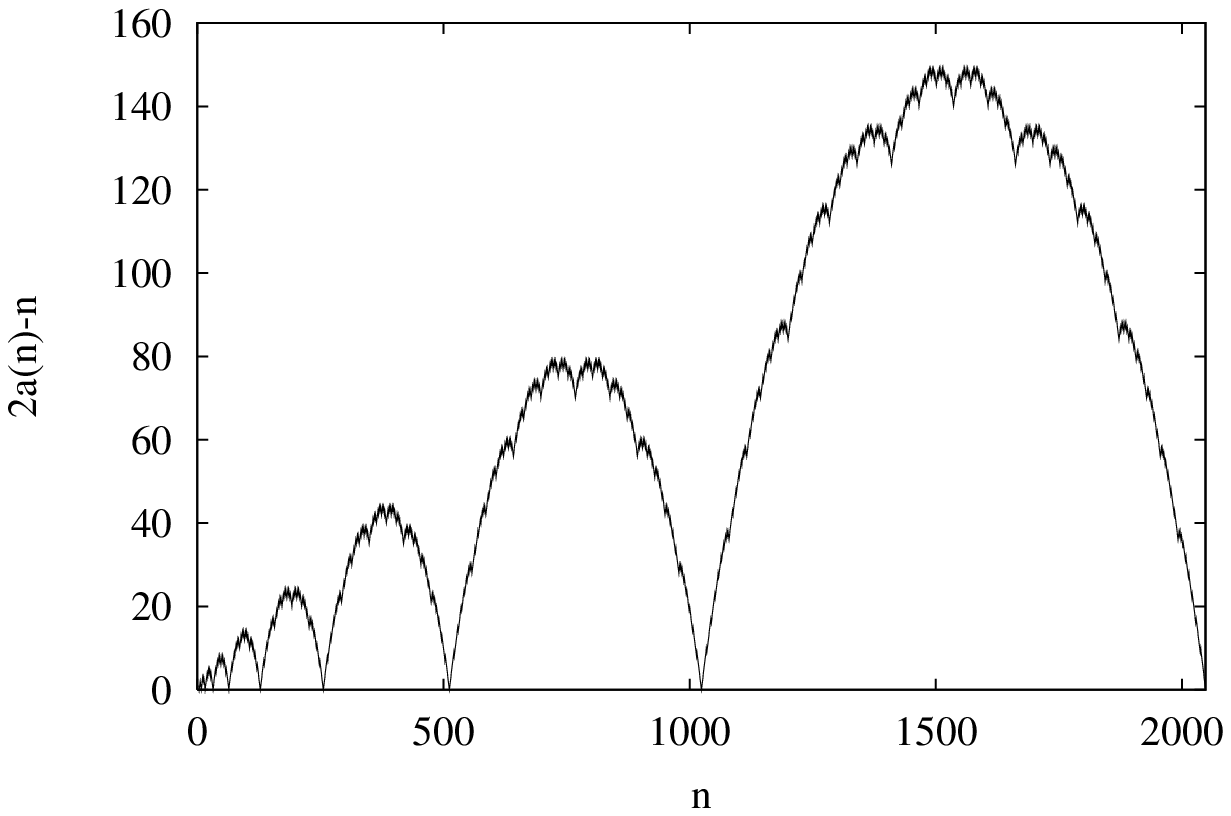}

\includegraphics[height=8cm]{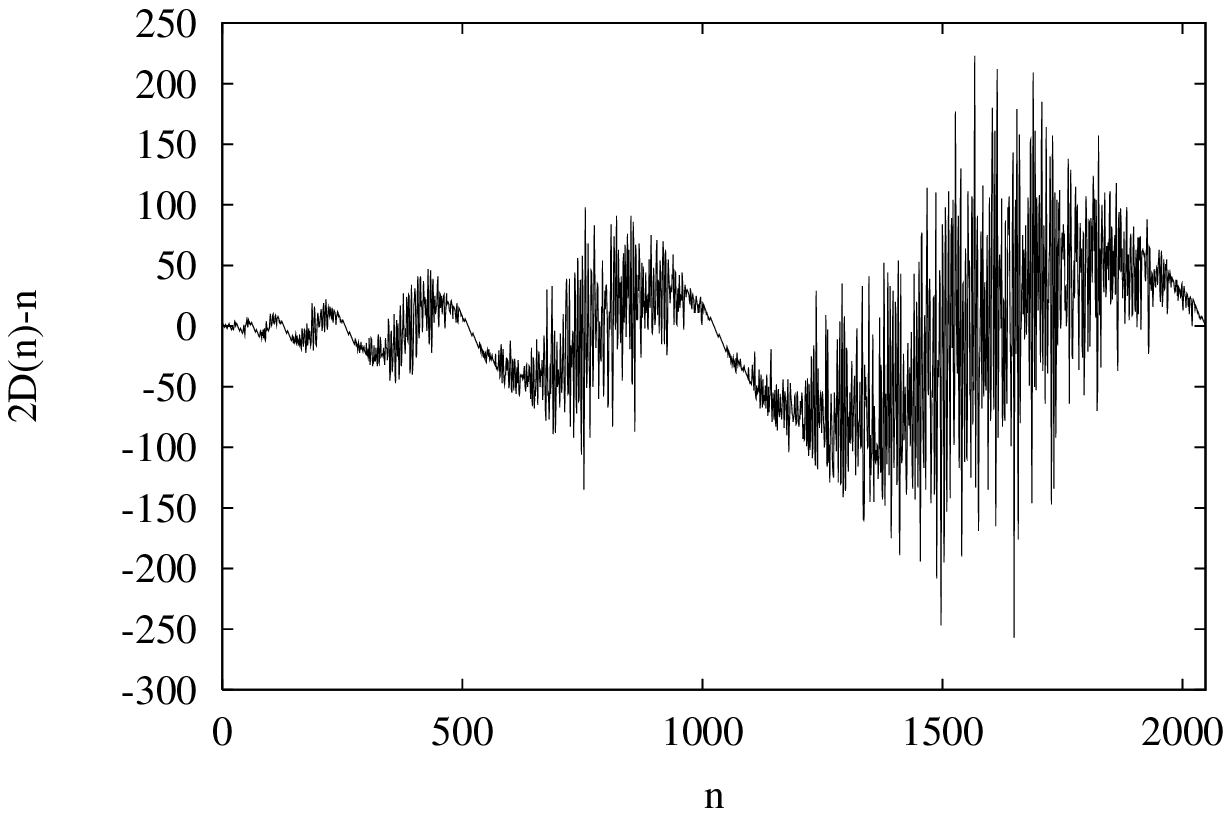}

\parbox[t]{.85\textwidth}
 {
 \caption[appear2]
 {\label{appear2}
\small
 Graphs of $2\, a(n) - n $ and $2 \, D(n)-n$.
}}
\end{center}
\end{figure}

Figure \ref{appear1} shows the first 2048 members of the $a$- and
$D$-sequences.  Both $a$ and $D$ are organized in ``generations'' of
increasing length and stay in some neighbourhood of $n/2$. These facts
become even more obvious when looking at $2\, a(n)-n$ and $2\,
D(n)-n$, see figure \ref{appear2}.

In order to make the ``genealogy'' more precise, we 
define a generation number $g(n)$ for each $n \geq 1$ through 
\be 
g(n) = \left\{
\begin{array}{ll}
0 &\mbox{\ if \ } n=1  \, , \\ 
k &\mbox{\ if \ } 2^{k-1}  < n \leq 2^k \mbox{\ \ for \ \ } n>1 \, . 
\end{array}
\right. 
\ee 
As in \cite{mypaper}, we interpret $D(n)$ as the 
sum of its {\em mother} at spot $D(n_1)$ and its {\em father} 
at $D(n_2)$, with 
\be 
\begin{array}{ll}
& n_1 = D(n-1) \, , \\ 
& n_2 = n-1-D(n-2) \, .  
\end{array}
\ee 
Tables \ref{parD} and \ref{parD1} show the structure of the
generations and the genealogy.  Inspecting an extended version of
table \ref{parD}, we make a number of observations, valid for
generation $k$, with $k\geq 5$: 

\begin{table}
\small 
\begin{center}
\begin{tabular}{ccccccc}
$k$ & $n$  & $n_1$ & $g(n_1)$ & $n_2$ & $g(n_2)$ & $D(n)$ \\
\hline 
\multicolumn{6}{c}{ } \\
{\bf 2} &    3   &     1 &  0    &    1 &  0    &    2 \\  
\hline 
&    4   &    2 &  1    &    2 &  1    &    2 \\[4mm]
{\bf 3}&    5   &     2 &  1    &    2 &  1    &    2 \\
&    6   &    2 &  1    &    3 &  2    &    3 \\
\hline 
&    7   &     3 &  2    &    4 &  2    &    4 \\
&    8   &     4 &  2    &    4 &  2    &    4 \\[4mm]       
{\bf 4} &    9   &     4 &  2    &    4 &  2    &    4 \\
&   10   &     4 &  2    &    5 &  3    &    4 \\
&   11   &     4 &  2    &    6 &  3    &    5 \\
\hline
&   12   &     5 &  3    &    7 &  3    &    6 \\
&   13   &     6 &  3    &    7 &  3    &    7 \\
&   14   &     7 &  3    &    7 &  3    &    8 \\
&   15   &     8 &  3    &    7 &  3    &    8 \\
&   16   &     8 &  3    &    7 &  3    &    8 \\[4mm]
{\bf 5} &   17   &     8 &  3    &    8 &  3    &   8 \\
&   18   &     8 &  3    &    9 &  4    &    8 \\
&   19   &     8 &  3    &   10 &  4    &    8 \\
&   20   &     8 &  3    &   11 &  4    &    9 \\
\hline 
&   21   &     9 &  4    &   12 &  4    &   10 \\
&   22   &    10 &  4    &   12 &  4    &   10 \\
&   23   &    10 &  4    &   12 &  4    &   10 \\
&   24   &    10 &  4    &   13 &  4    &   11 \\
&   25   &    11 &  4    &   14 &  4    &   13 \\
&   26   &    13 &  4    &   14 &  4    &   15 \\
&   27   &    15 &  4    &   13 &  4    &   15 \\
&   28   &    15 &  4    &   12 &  4    &   14 \\
\hline 
&   29   &    14 &  4    &   13 &  4    &   15 \\
&   30   &    15 &  4    &   15 &  4    &   16 \\
&   31   &    16 &  4    &   15 &  4    &   16 \\
&   32   &    16 &  4    &   15 &  4    &   16 \\               
 \end{tabular}
\parbox[t]{.85\textwidth}
 {
 \caption[parD]
 {\label{parD}
\small
Genealogy in the $D$-sequence.
Head, body and tail of a generation are separated by horizontal lines.
Note that tails are not defined for $k < 5$. 
}}
\end{center}
\end{table}

\begin{table}
\small 
\begin{center}
\begin{tabular}{ccccccc}
$k$ & $n$  & $n_1$ & $g(n_1)$ & $n_2$ & $g(n_2)$ & $D(n)$ \\
\hline 
\multicolumn{6}{c}{ } \\
{\bf 6} &    33   &     16 &   4   &     16 &   4   &     16 \\
 &   34  &      16  &  4    &    17  &  5   &     16 \\
 &   35  &      16  &  4    &    18  &  5   &     16 \\
 &   36  &      16  &  4    &    19  &  5   &     16 \\
 &   37  &      16  &  4    &    20  &  5   &     17 \\
\hline 
 &   38  &      17  &  5    &    21  &  5   &     18 \\
 &   39  &      18  &  5    &    21  &  5   &     18 \\
 &   40  &      18  &  5    &    21  &  5   &     18 \\
 &   41  &      18  &  5    &    22  &  5   &     18 \\
 &   42  &      18  &  5    &    23  &  5   &     18 \\
 &   43  &      18  &  5    &    24  &  5   &     19 \\
 &   44  &      19  &  5    &    25  &  5   &     21 \\
 &   45  &      21  &  5    &    25  &  5   &     23 \\
 &   46  &      23  &  5    &    24  &  5   &     21 \\
 &   47  &      21  &  5    &    23  &  5   &     20 \\
 &   48  &      20  &  5    &    26  &  5   &     24 \\
 &   49  &      24  &  5    &    28  &  5   &     25 \\
 &   50  &      25  &  5    &    25  &  5   &     26 \\
 &   51  &      26  &  5    &    25  &  5   &     28 \\
 &   52  &      28  &  5    &    25  &  5   &     27 \\
 &   53  &      27  &  5    &    24  &  5   &     26 \\
 &   54  &      26  &  5    &    26  &  5   &     30 \\
 &   55  &      30  &  5    &    28  &  5   &     30 \\
 &   56  &      30  &  5    &    25  &  5   &     29 \\
 &   57  &      29  &  5    &    26  &  5   &     30 \\
 &   58  &      30  &  5    &    28  &  5   &     30 \\
 &   59  &      30  &  5    &    28  &  5   &     30 \\
\hline 
 &   60  &      30  &  5    &    29  &  5   &     31 \\
 &   61  &      31  &  5    &    30  &  5   &     32 \\
 &   62  &      32  &  5    &    30  &  5   &     32 \\
 &   63  &      32  &  5    &    30  &  5   &     32 \\
 &   64  &      32  &  5    &    31  &  5   &     32 \\
 \end{tabular}
\parbox[t]{.85\textwidth}
 {
 \caption[parD1]
 {\label{parD1}
\small
Continuation of table \ref{parD}. 
}}
\end{center}
\end{table}

\begin{itemize} 
\item[C1] For the first $k-2$ members the function
      $D$ takes the value $2^{k-2}$. The ($k-1$)th element 
      is $2^{k-2}+1$. (The first $k-1$ members of a generation 
      will be called {\em head}.) 
\item[C2] For the last $k-2$ members the function
      $D$ takes the value $2^{k-1}$. The element just before the 
      last $k-2$ members takes the value $2^{k-1}-1$.
      (The last $k-1$ members will be called {\em tail}.)
\item[C3] The last member of generation $k-2$ is simultaneously 
      the mother of all head members and the father of the
      first head member. 
      The fathers of the remaining head members are (in ascending order) 
      the members of the head of generation $k-1$.  
\item[C4] The parents of tail members are tail members 
          of generation $k-1$.
\item[C5] The values of $D(n)$ lie in the range $[2^{k-2},2^{k-1}]$. 
\end{itemize}

\begin{figure}
\begin{center}
\includegraphics[height=7cm]{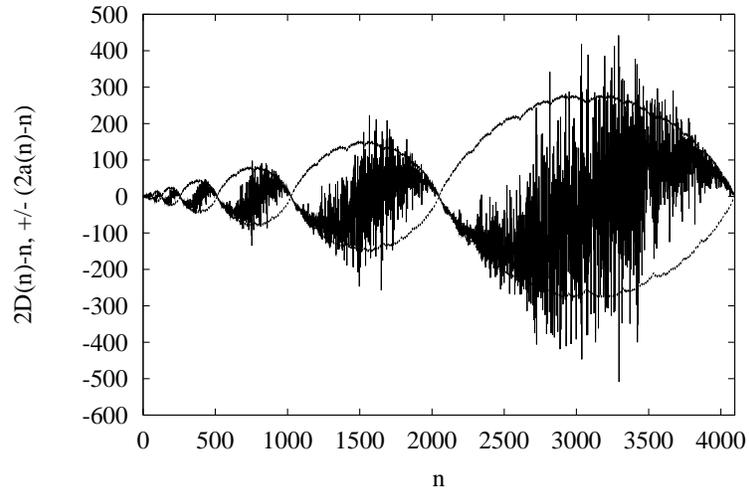}
\parbox[t]{.85\textwidth}
 {
 \caption[Dahull]
 {\label{Dahull}
\small
 Graph of $2\, D(n)-n$, together with $\pm(2 \, a(n)-n)$.
}}
\end{center}
\end{figure}

An interesting observation can be made when one plots together $D(n)$
and $a(n)$.  Figure \ref{Dahull} shows $2\, D(n)-n$, together with
$\pm (2\, a(n)-n)$.  The latter two functions nicely model the
``outer'' boundary of the fluctuating $D(n)$ in some neighbourhood of
the generation boundaries.

\begin{figure}
\begin{center}
\includegraphics[width=6cm]{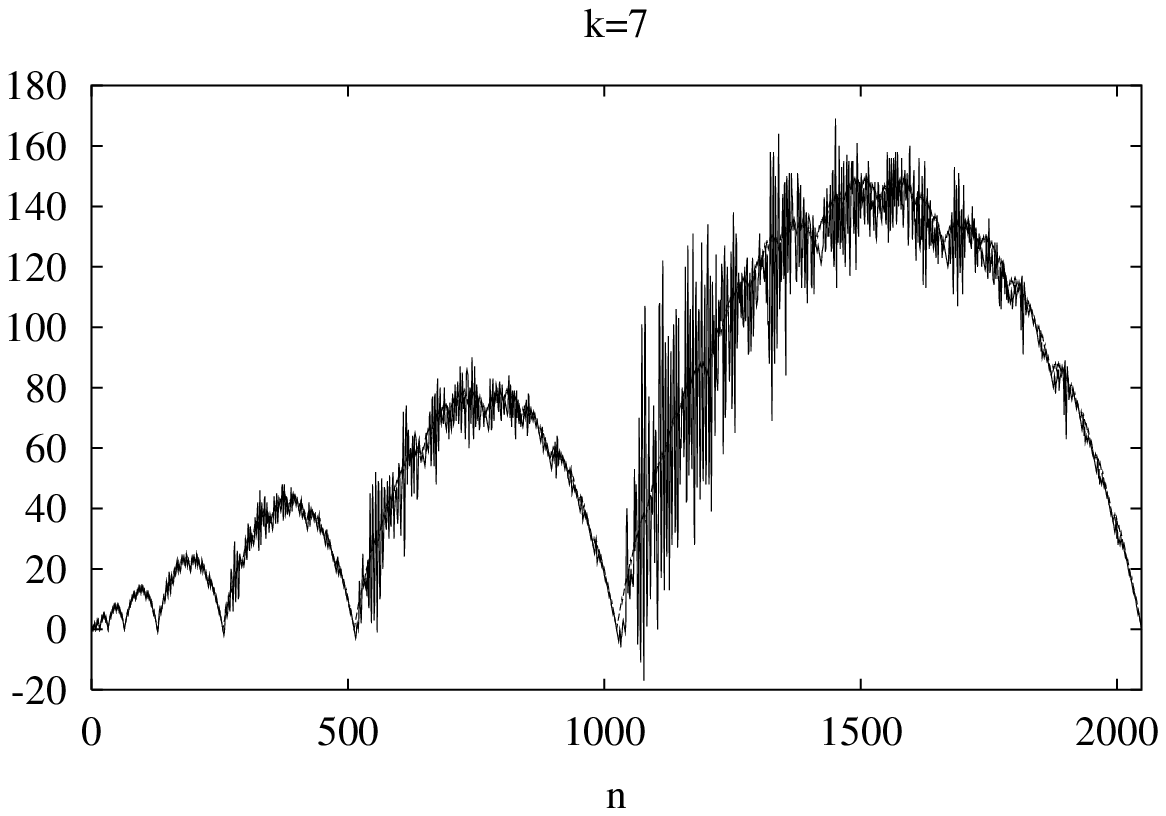}
\includegraphics[width=6cm]{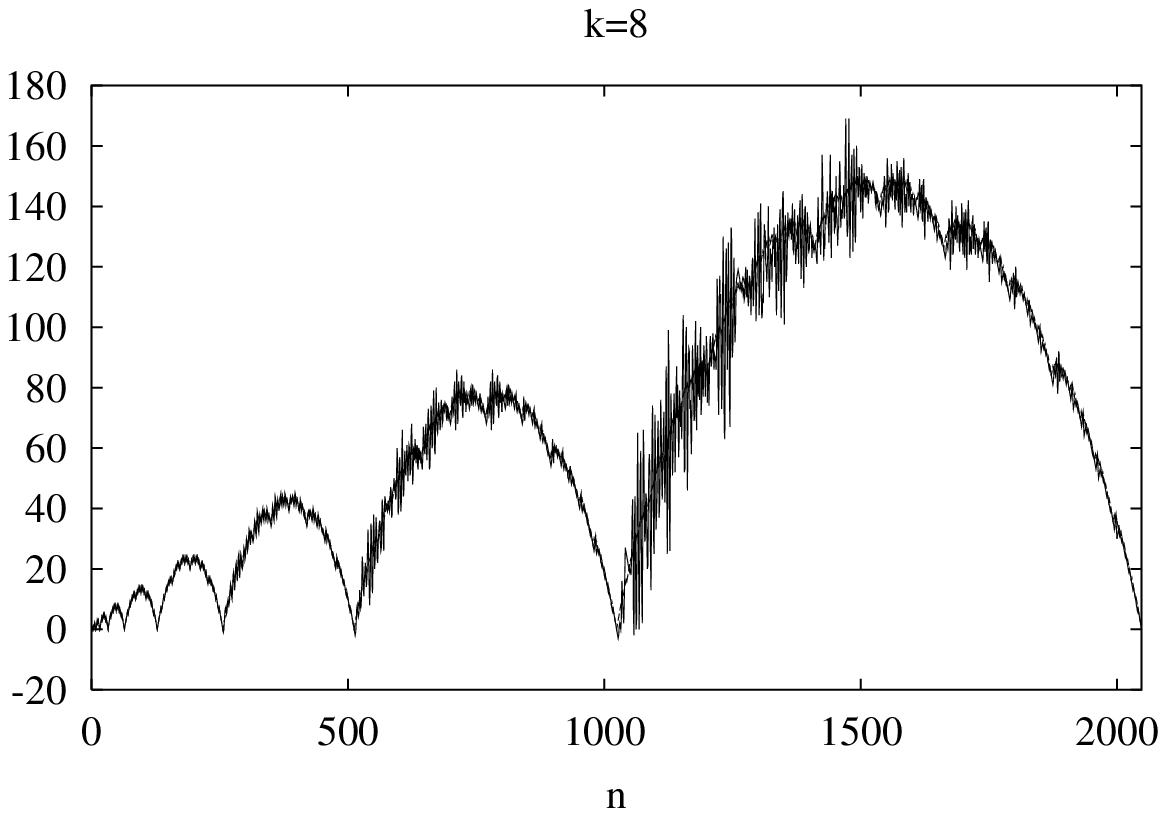}

\includegraphics[width=6cm]{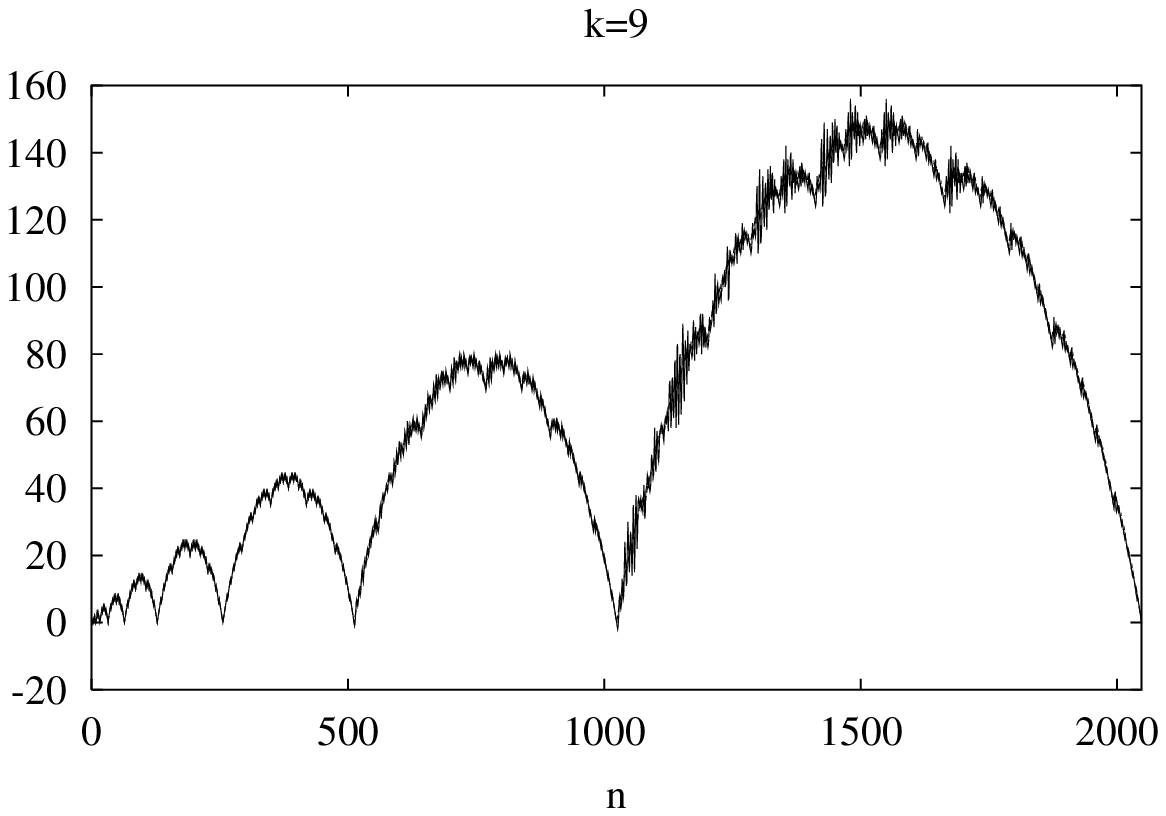}
\includegraphics[width=6cm]{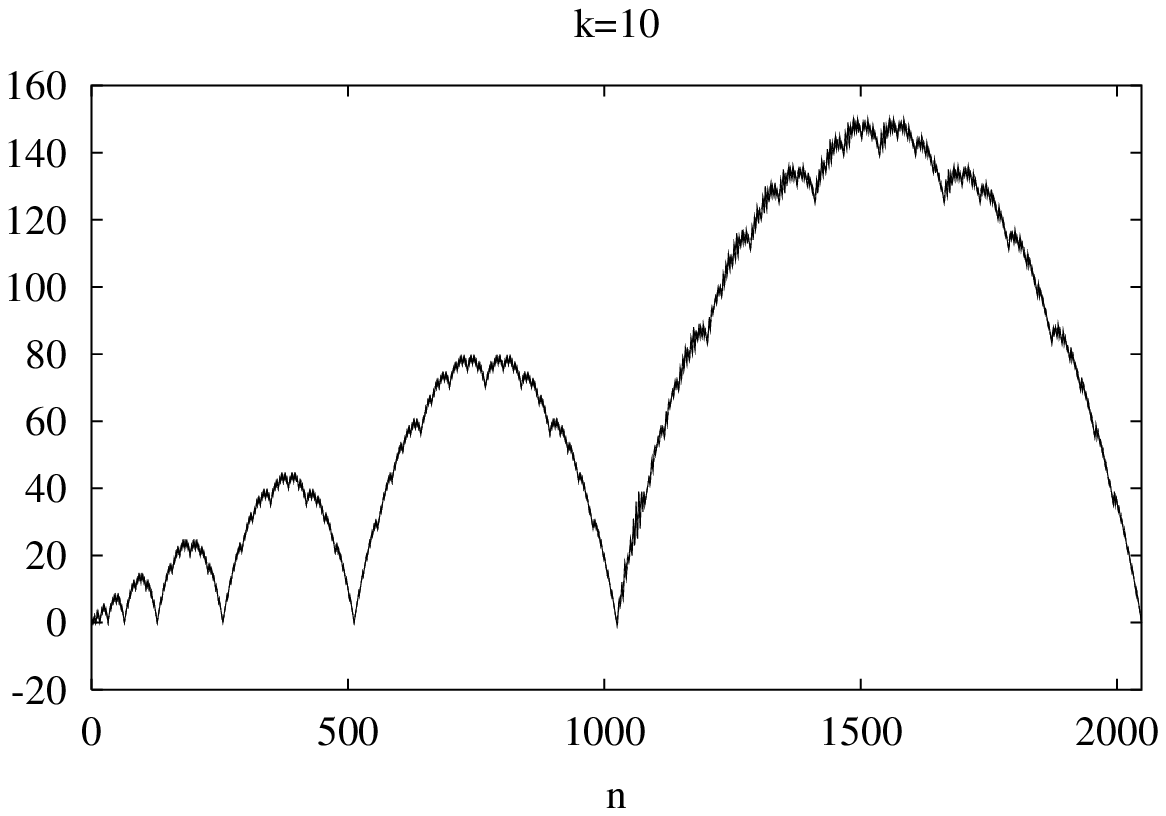}

\parbox[t]{.85\textwidth}
 {
 \caption[hybrid]
 {\label{hybrid}
\small
Graphs of $2 \, aD_k(n) - n$, for $k=7$, 8, 9, 10. 
}}
\end{center}
\end{figure}

The close relation of $a$ and $D$ is also underlined by the 
following experiment: Use the $a$-recurrence to 
generate the first $k$ generations of numbers. Then continue 
with the recursion relation of the $D$-numbers. 
Graphs illustrating the behaviour of the resulting function, 
to be called $aD_k(n)$, are
shown in figure \ref{hybrid}, for $k$ from 7 to 10. 
Plotted are $2\, aD_k(n) - n$, together with $2 \, a(n)-n$.
With increasing $k$ the ``chaotic'' fluctuations get reduced and the 
function becomes very similar to $a(n)$. It seems that one 
can in this way generate a large family of sequences with 
different ``levels of chaos''.

\section{Statistical Properties} 

\begin{figure}
\begin{center}
\includegraphics[width=6cm]{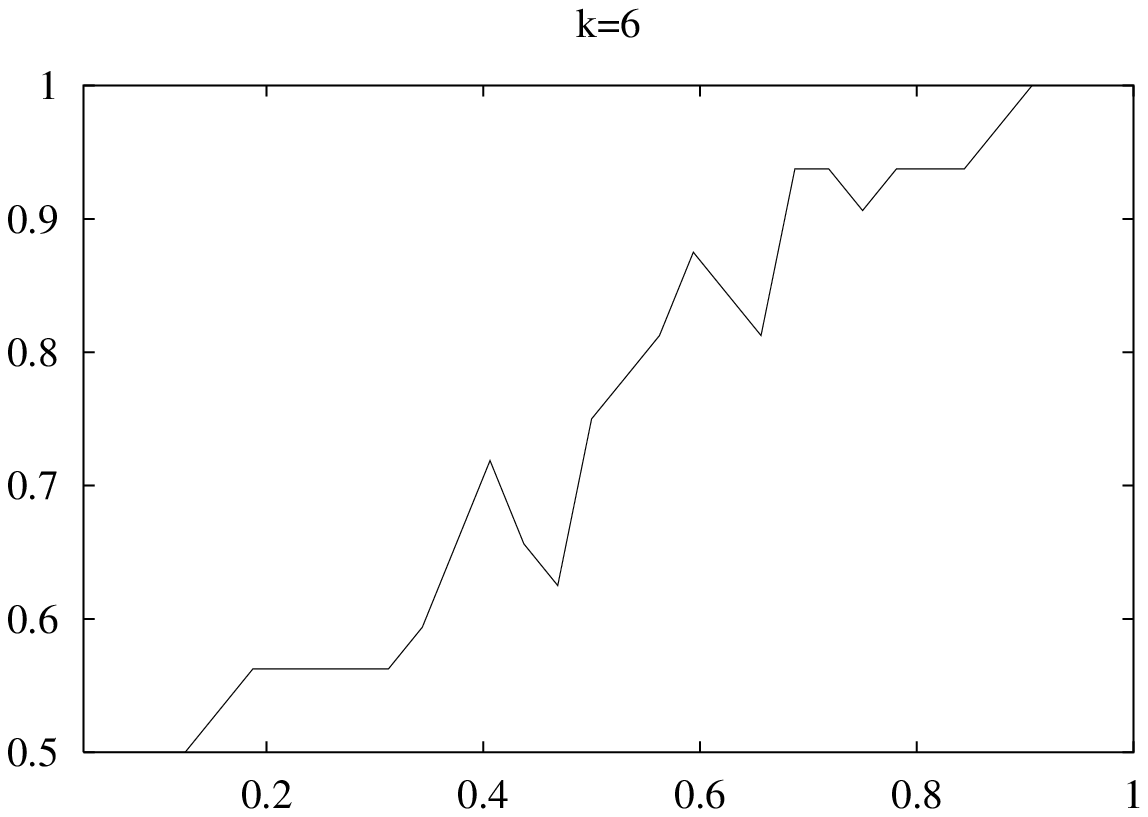}
\includegraphics[width=6cm]{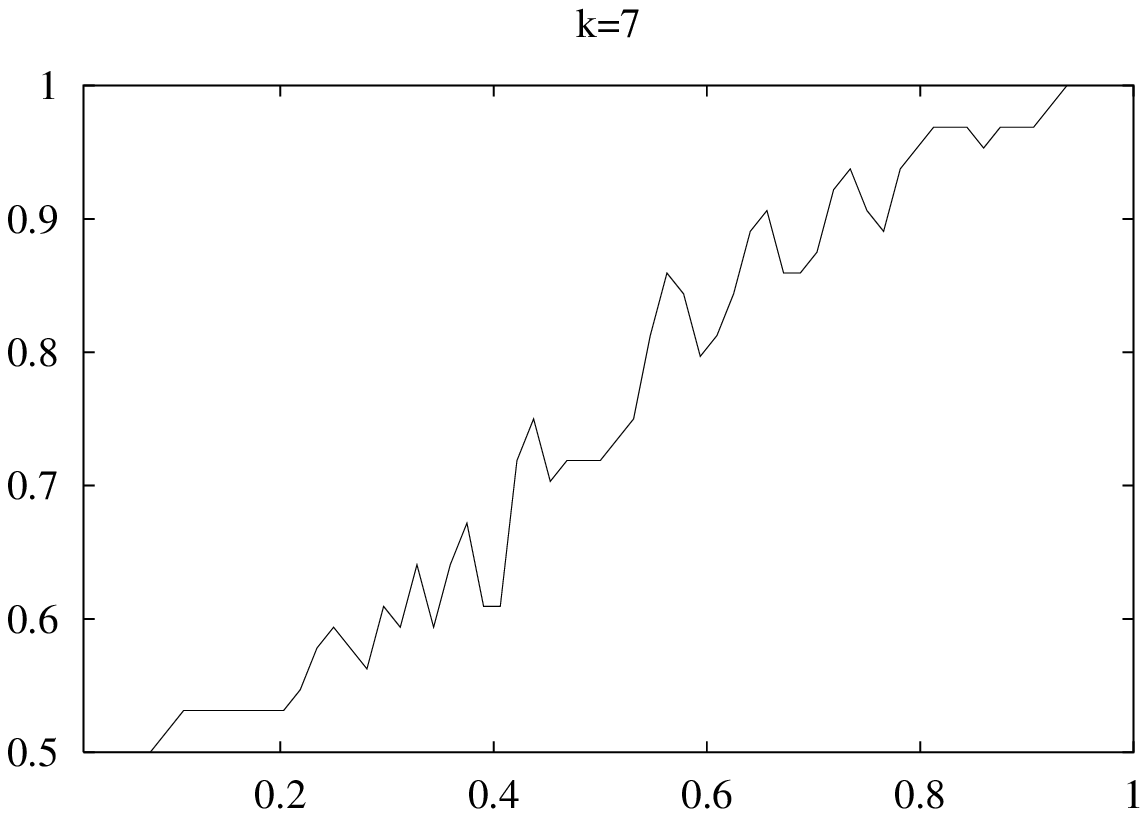}

\includegraphics[width=6cm]{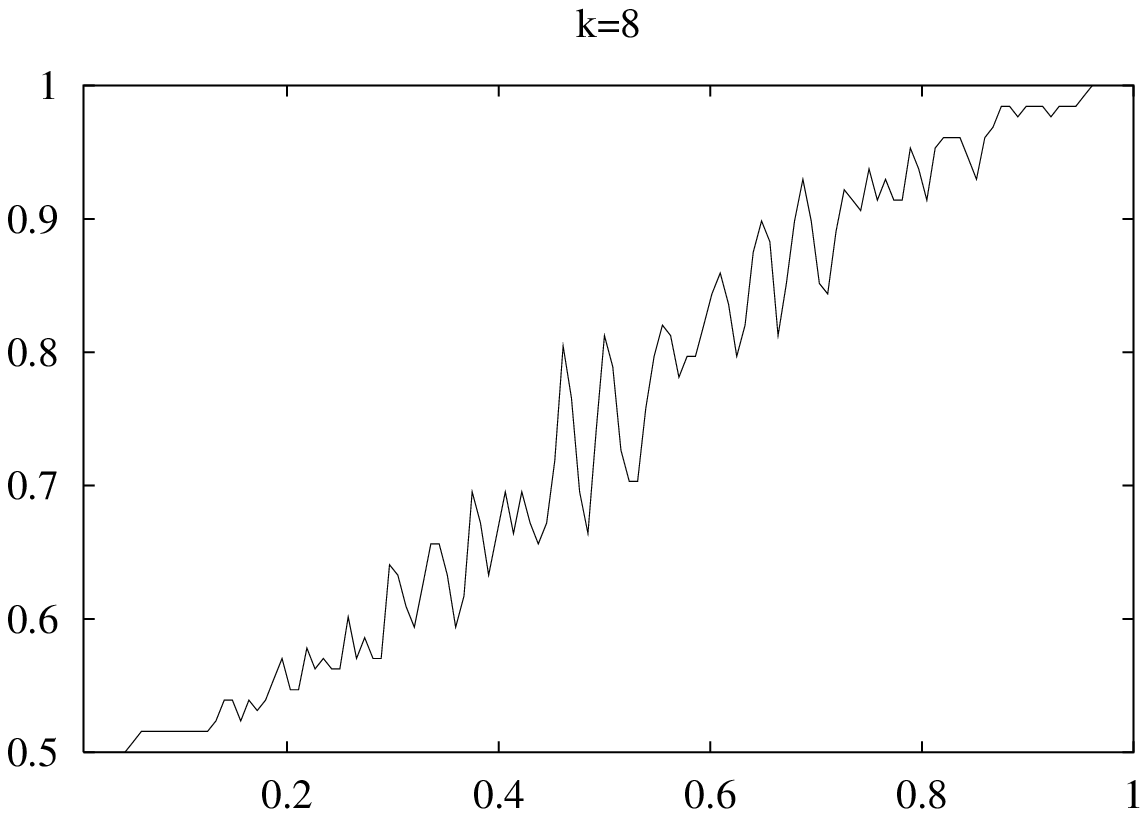}
\includegraphics[width=6cm]{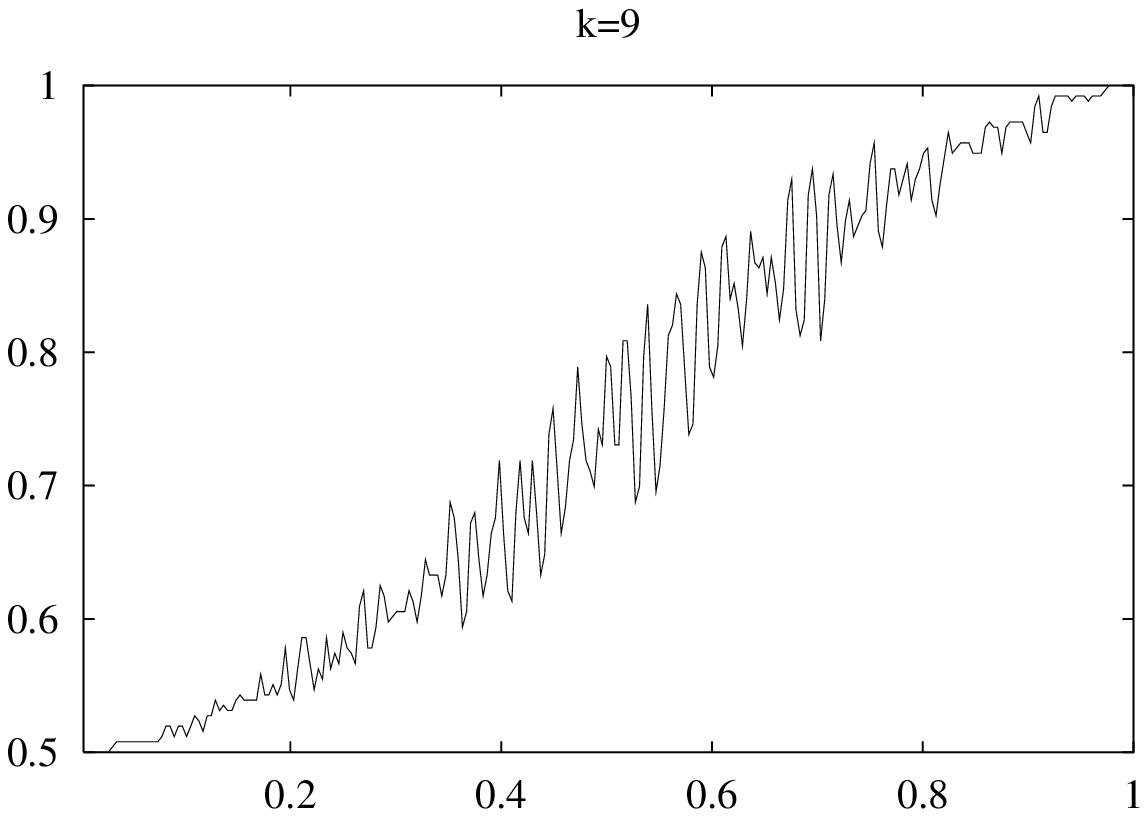}

\includegraphics[width=6cm]{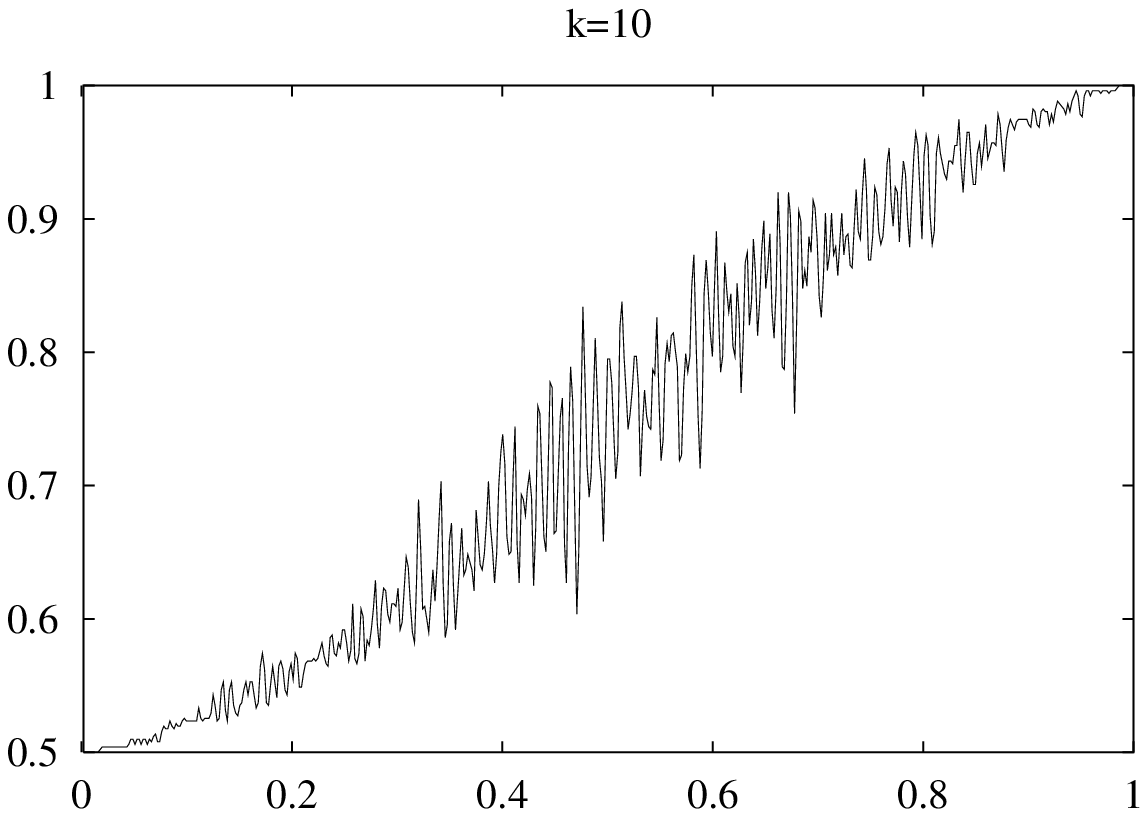}
\includegraphics[width=6cm]{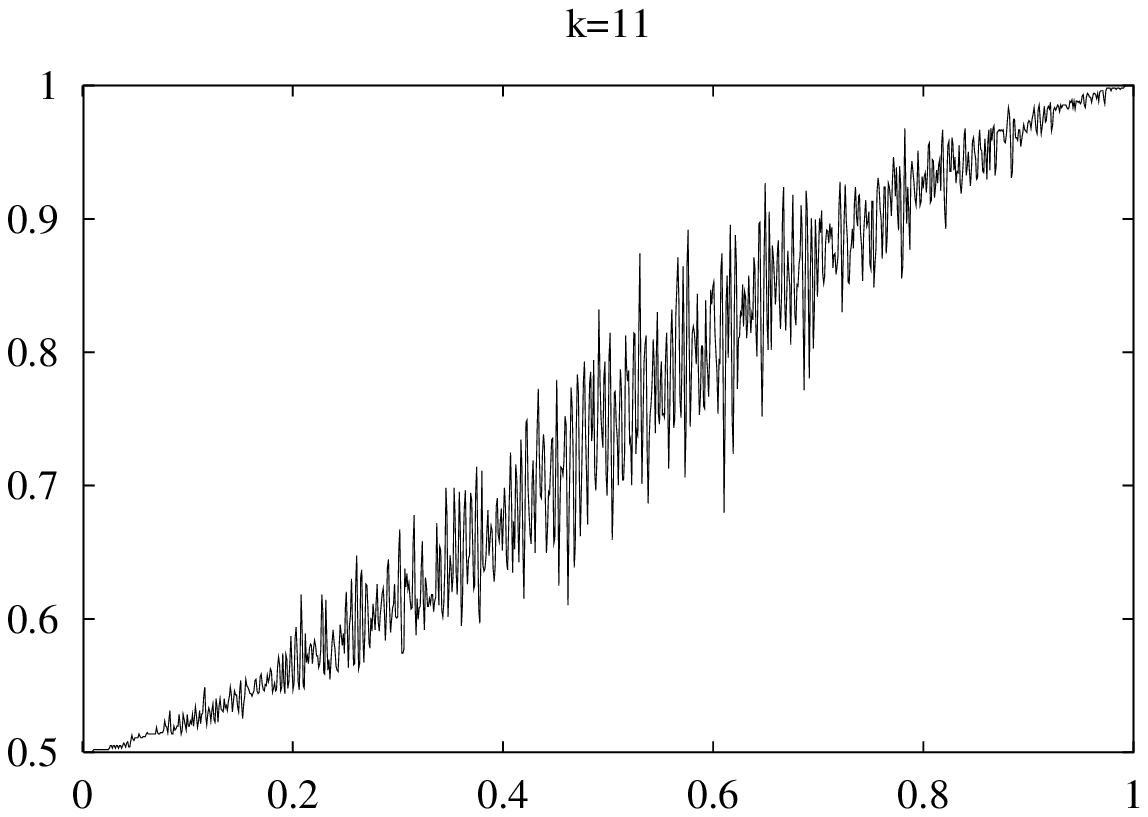}

\includegraphics[width=6cm]{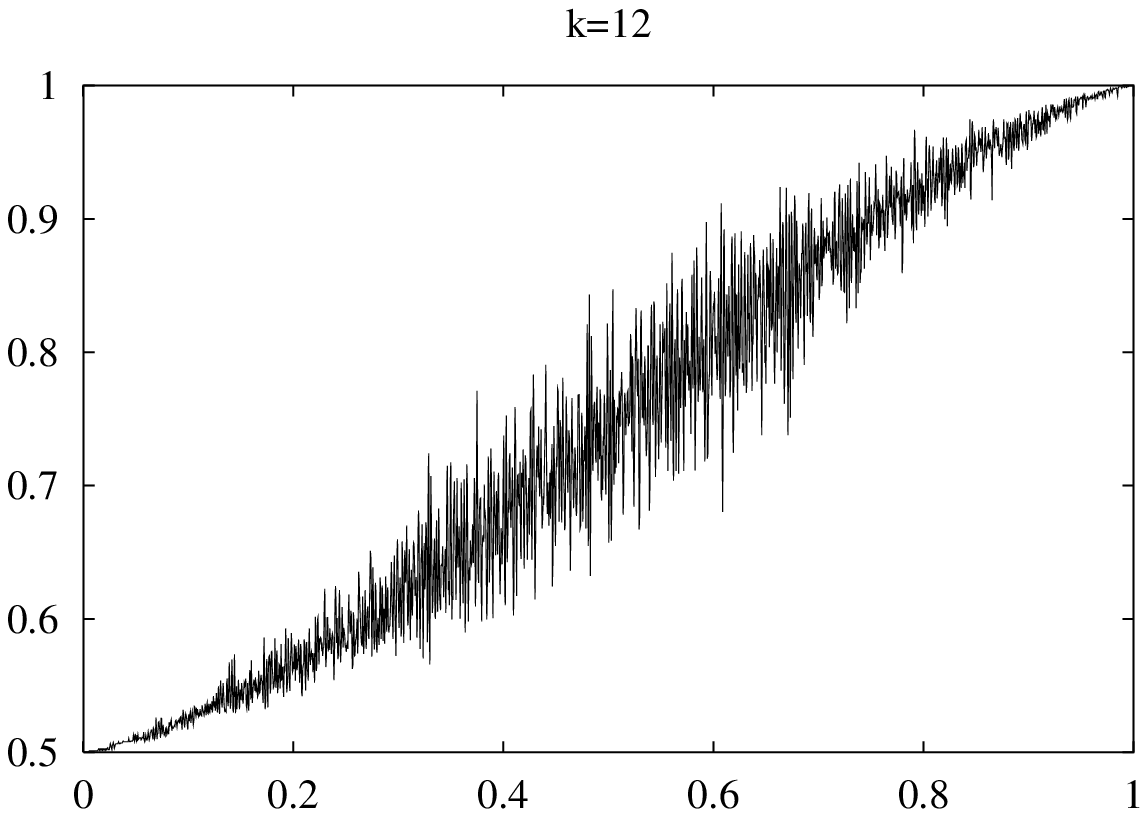}
\includegraphics[width=6cm]{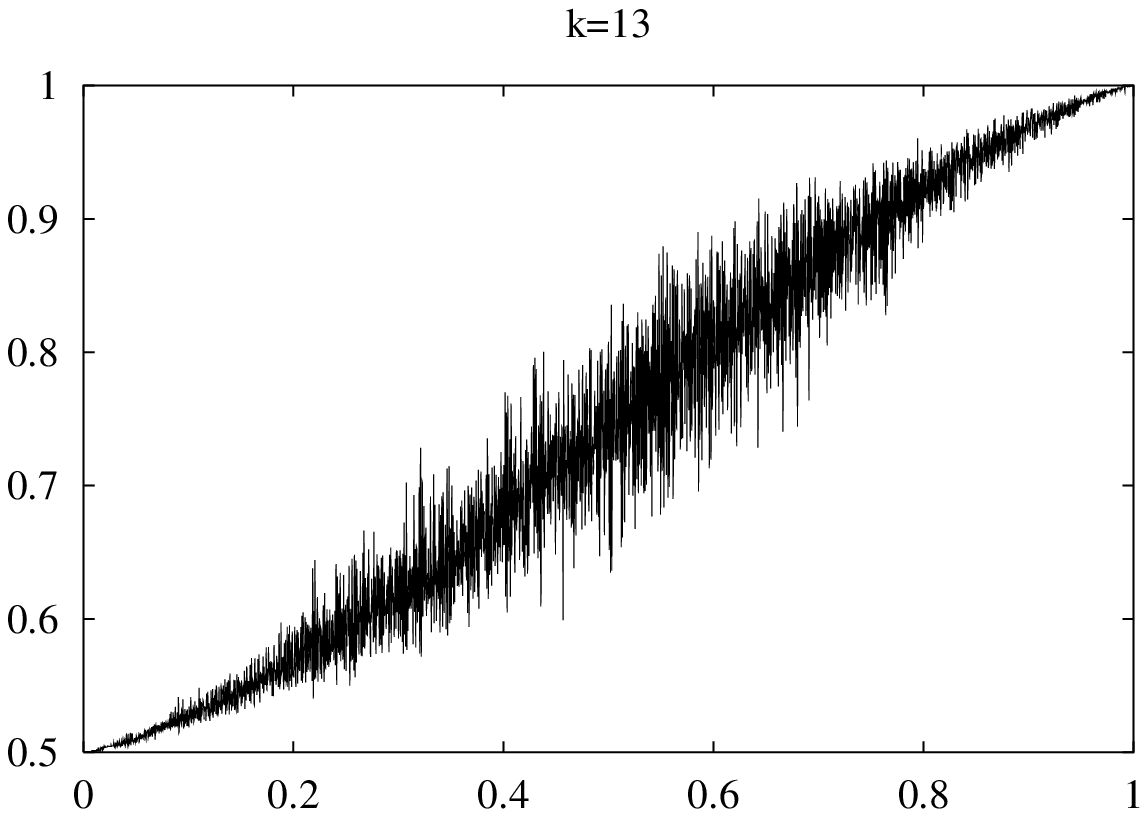}

\parbox[t]{.85\textwidth}
 {
 \caption[boxes]
 {\label{boxes}
\small
 Rescaled graphs of $D(n)$, for generations $6 \leq k \leq 13$. 
}}
\end{center}
\end{figure}

In generation $k$, i.e.\ 
for $2^{k-1} < n \leq 2^k$, the function $D(n)$ takes values 
in the range $2^{k-2} < n \leq 2^{k-1}$.
It seems natural to plot $y = D(n)/2^{k-1}$ in terms of 
$x = (n - 2^{k-1})/2^{k-1}$. We have $0 < x \leq 1$, and 
$y \leq 0.5 \leq 1$. 
Plots of this type for generations 6 to 13 are shown 
in figure~\ref{boxes}. The similarity of the graphs 
suggests that there could be some statistical properties  
becoming independent of $k$ when $k$ becomes large.

\subsection{Step Size Statistics}

The function inside a given generation may be considered representing
a random walk of $2^{k-1}-1$ steps, starting from $2^{k-2}$ and arriving
at $2^{k-1}$. It is interesting to look at the distribution of 
the step sizes. Let us define 
\be 
S(n)= D(n)-D(n-1) \, . 
\ee 
The square of the variance of this quantity is given by  
\be 
M(k)^2 =
\bigl\langle
S(n)^2
\bigr\rangle_k
-
\bigl\langle
S(n)
\bigr\rangle_k^2 \, ,
\ee 
where $\langle (.) \rangle_k$ denotes the average over
the $k$-th generation.
Table~\ref{msqr} shows numerical results for $ \ln_2 M(k)$
for generations 13 to 25 and also the logarithmic ratios
$ \alpha_k = \ln_2 (M(k)/M(k-1))$.                                      
$\ln_2$ denotes the logarithm with respect to base 2.
The results for the latter quantity converge to 0.88(1). 
We conclude that
\be 
\frac{M(k)}{M(k-1)} \simeq 2^{\, \alpha} \, ,
\ee 
with $\alpha = 0.88(1)$. This exponent is consistent with the 
one found for the Hofstadter $Q(n)$ \cite{mypaper}.

\begin{table}
\small 
\begin{center}
\begin{tabular}{c|cc}
$k$ & $\ln_2 M(k)$ &  $\alpha_k$  \\
\hline
  13   &  \phantom{1}6.857    &       0.949  \\
  15   &  \phantom{2}8.683    &       0.910  \\
  17   &    10.498    &       0.896  \\
  19   &    12.291    &       0.888  \\
  21   &    14.071    &       0.888  \\
  22   &    14.961    &       0.890  \\
  23   &    15.845    &       0.884  \\
  24   &    16.726    &       0.882  \\
  25   &    17.598    &       0.872  \\
 \hline
 \end{tabular}
\parbox[t]{.85\textwidth}
 {
 \caption[msqr]
 {\label{msqr}
\small
Variances $M(k)$ and ratios $\alpha_k = \ln_2 (M(k)/M(k-1))$. \\
}}
\end{center}
\end{table}
     
\begin{figure}
\begin{center}
\includegraphics[height=7cm]{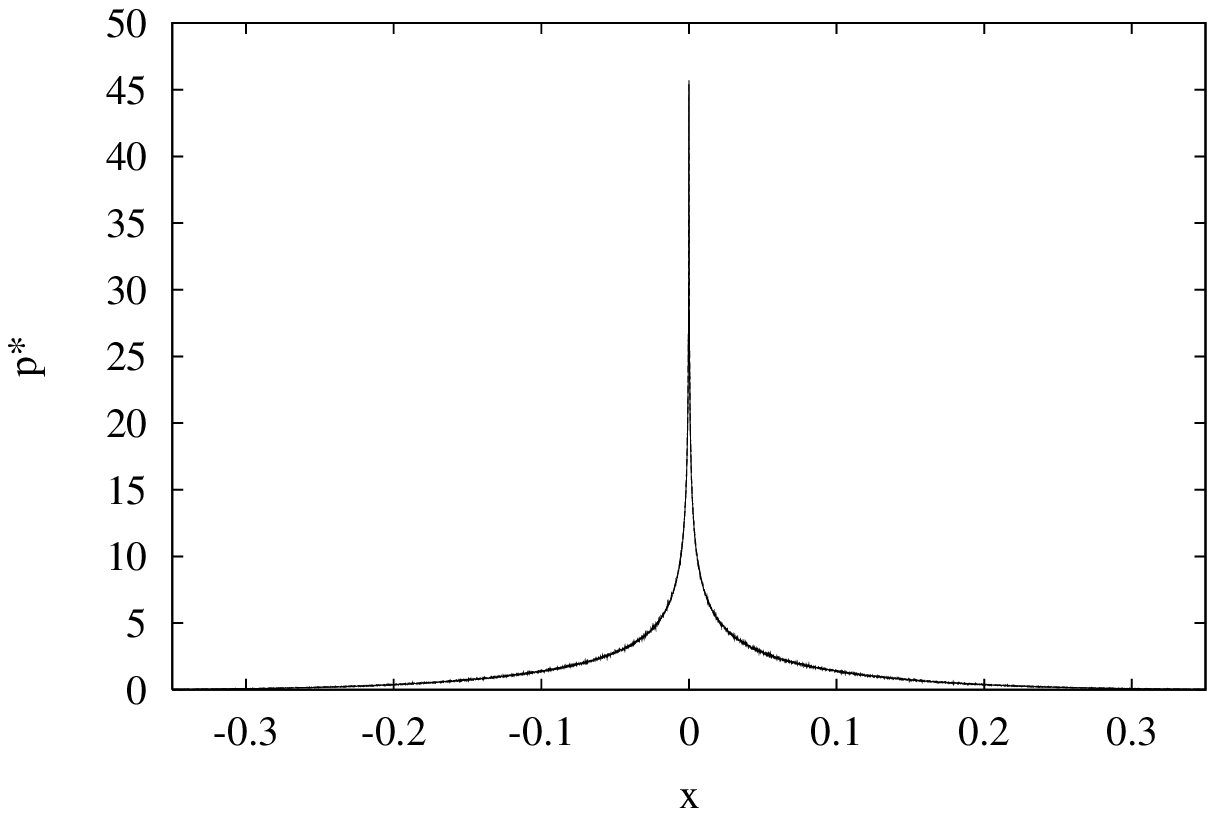}

\includegraphics[height=7cm]{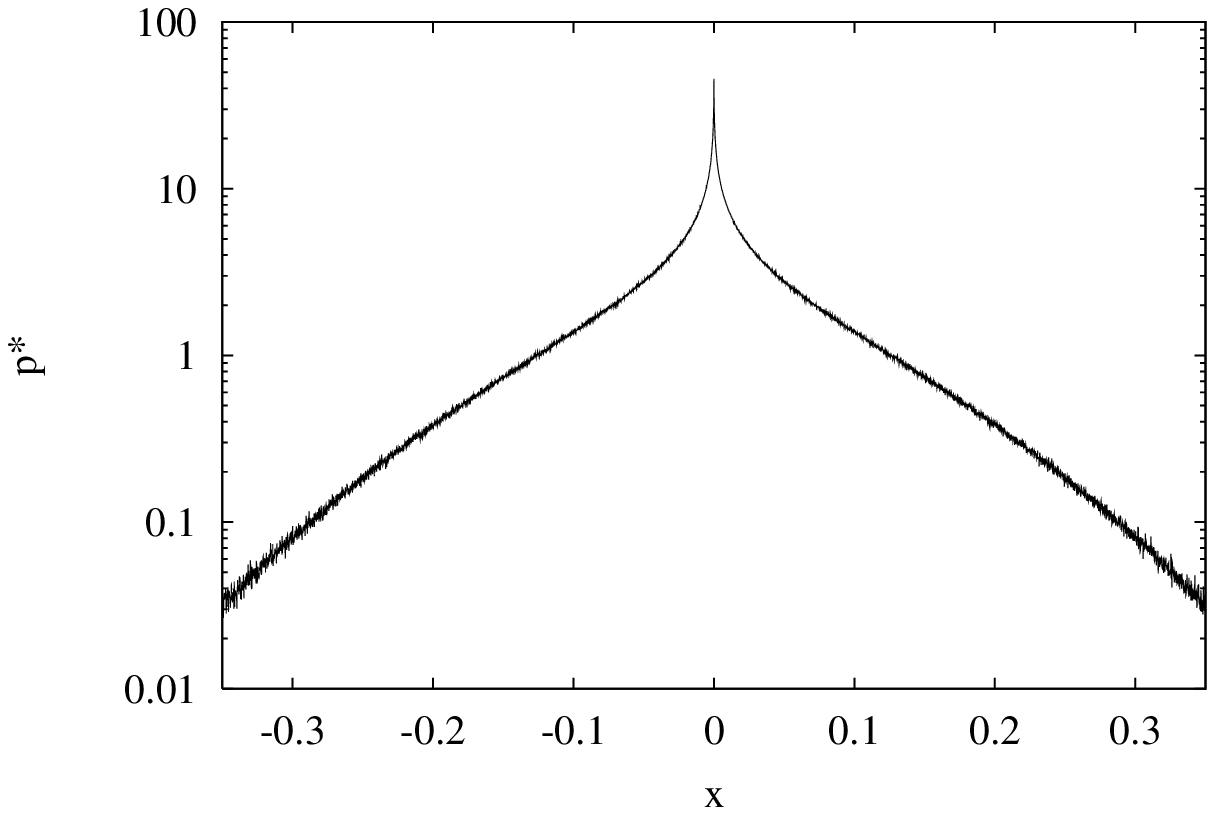}

\parbox[t]{.85\textwidth}
 {
 \caption[pdist]
 {\label{pdist}
\small
 Statistical distribution of $x= (D(n)-D(n-1))/2^{0.88\,(k-1)}$, 
 in generations $k=24$ and $k=25$ (top). Same distribution on 
 logarithmic scale for $k=25$ (bottom). 
}}
\end{center}
\end{figure}

Figure \ref{pdist} shows a histogram $p^*$ of the 
variable $x= S(n)/2^{0.88 \, (k-1)}$, for $k= 24$ and $k=25$, 
plotted on top of each other. The two histograms match nicely. 
The statistical distribution for $k=25$ is plotted on a logarithmic scale 
in the lower part of the figure.  As was the case with the distribution
function of suitable $Q$-number observables, the tails
can be nicely fitted with a properly rescaled error function erfc,
defined through
\be 
{\rm erfc}(x) = \frac{2}{\sqrt{\pi}}
\int_x^\infty dt \, \exp(-t^2) \, .      
\ee 

\begin{figure}
\begin{center}
\includegraphics[height=8cm]{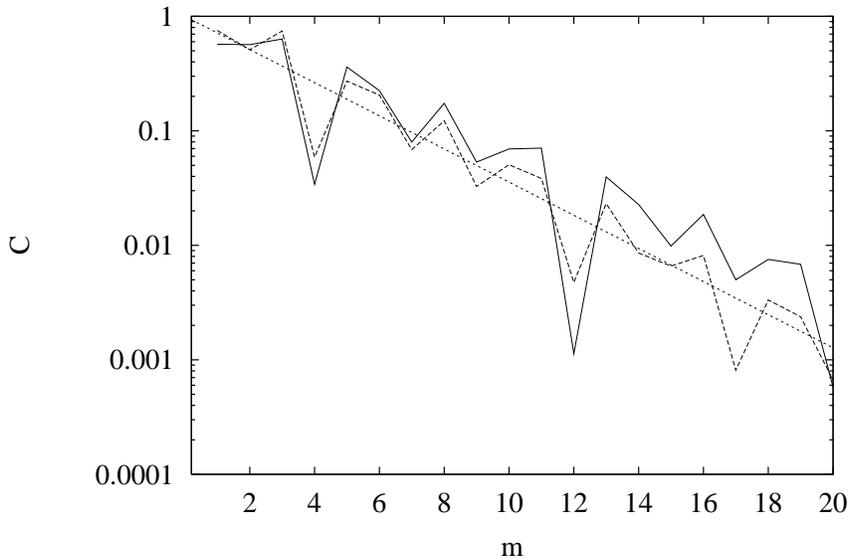}
\parbox[t]{.85\textwidth}
 {
 \caption[lamdec]
 {\label{lamdec}
\small
$C_D = |\lambda_m^2 - 1.57|$ (full lines),  
$C_Q = |\lambda_m^2 - 2|$ (dotted lines), and 
the function $\exp(-m/3)$.
}}
\end{center}
\end{figure}

In \cite{mypaper} it was observed that the probability density 
$p_m(x_m)$ of the rescaled difference 
$x_m = (Q(n)-Q(n-m))/n^\alpha$ was (up to a rescaling)
with high precision identical with the distribution 
$p^*$ of $(Q(n)-n/2)/n^\alpha$, i.e., 
\be \label{scale}
p_m(x_m) = \lambda_m p^*(x_m/\lambda_m) \, . 
\ee
A similar type of scaling applies here.
We define 
\be 
x_m = (D(n)-D(n-m))/2^{\alpha \, (k-1)} \, . 
\ee 
Note that in the present case $p^*$ is the distribution of 
$x_1$. One observes validity of eq.~(\ref{scale}) with very 
good precision for $m \geq 2$. 
One can determine the $\lambda_m$ from the second moments,  
\be 
\lambda_m^2 = \frac{ \langle x_m^2 \rangle - \langle x_m \rangle^2 }
                   { \langle x_1^2 \rangle - \langle x_1 \rangle^2 } \, . 
\ee 
They 
converge against $\lambda_\infty^2 \approx 1.57$. Looking at 
\be
C = |\lambda_m^2 - \lambda_{\infty}^2|
\ee  
as function of $m$, we observe a striking similarity with 
the corresponding function for the Hofstadter sequence, 
see figure \ref{lamdec}. The ups and downs in  both cases 
are very similar. The decay goes like $\exp(-m/3)$.

\subsection{Numbers Left Out and Frequency Counting}

It was observed by A. K. Yao that the $Q$-sequence 
seems to leave out infinitely many numbers \cite{Yao}. 

\begin{figure}
\begin{center}
\includegraphics[height=7cm]{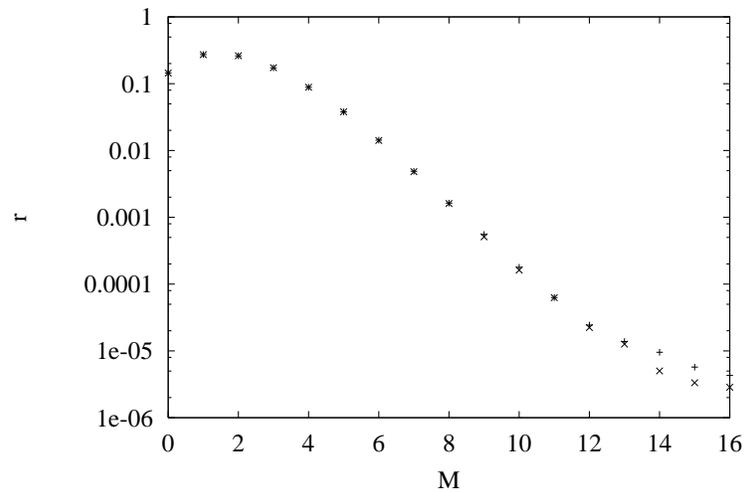}
\parbox[t]{.85\textwidth}
 {
 \caption[wieoft1]
 {\label{wieoft1}
\small
 $r(M)$, for $k=23$ (+) and $k=24$ (x). 
}}
\end{center}
\end{figure}

The $D$-function maps generation $k$, i.e.\ the range $[2^{k-1}+1,2^k]$, 
to the interval $I_k= [2^{k-2},2^{k-1}]$. We consider the question which 
fraction $r(M)$ of the $2^{k-2}+1$ 
numbers in $I_k$ are generated exactly $M$ times. 
It turns out that 
these fractions converge with increasing $k$. 
Table \ref{wieoft1t} shows $r(M)$, $M \leq 6$, for $k=23$ and $k=24$. 
The $D$-function omits some 14 \% of all numbers. 
The table also shows $r(M)$ 
for the sequences $Q$, $F_{10}$, and $F_{11}$. The latter two sequences 
are close relatives of the Hofstadter sequence and will be introduced 
in section 4. 
There is a  fair agreement of the ratios $r(M)$ for all the four
sequences. 
Figure \ref{wieoft1}
shows $r(M)$ for $M \leq 16$ in generations 23 and 24. 
Only a small deviations between the two sets of numbers is seen 
for larger $M$. 

\begin{table}
\begin{center}
\begin{tabular}{c|cc|c|c|c}
$M$ & $k=23$ & $k=24$  & $Q$, $n < 2^{21.5}$ & $F_{10}$, $n < 2^{22}$ 
& $F_{11}$, $n < 2^{22}$ \\
\hline
     0 &  0.1446 & 0.1443 &  0.1358 & 0.1358 & 0.1342  \\
     1 &  0.2728 & 0.2722 &  0.2709 & 0.2706 & 0.2697  \\
     2 &  0.2615 & 0.2624 &  0.2700 & 0.2703 & 0.2709  \\
     3 &  0.1730 & 0.1731 &  0.1803 & 0.1804 & 0.1810  \\
     4 &  0.0885 & 0.0886 &  0.0900 & 0.0903 & 0.0909  \\ 
     5 &  0.0379 & 0.0380 &  0.0362 & 0.0361 & 0.0365  \\ 
     6 &  0.0143 & 0.0141 &  0.0122 & 0.0120 & 0.0122  \\ 
 \hline
 \end{tabular}
\parbox[t]{.85\textwidth}
 {
 \caption[wieoft1t]
 {\label{wieoft1t}
\small
Relative frequency $r(M)$ of numbers in $I_k$ 
that are generated by $D$ exactly 
$M$ times. The last column gives estimates for the $r$-ratios of the 
sequences $Q$, $F_{10}$, and $F_{11}$. The latter two recurrences will
be introduced in section 5. 
}}
\end{center}
\end{table}

\begin{figure}
\begin{center}
\includegraphics[height=7cm]{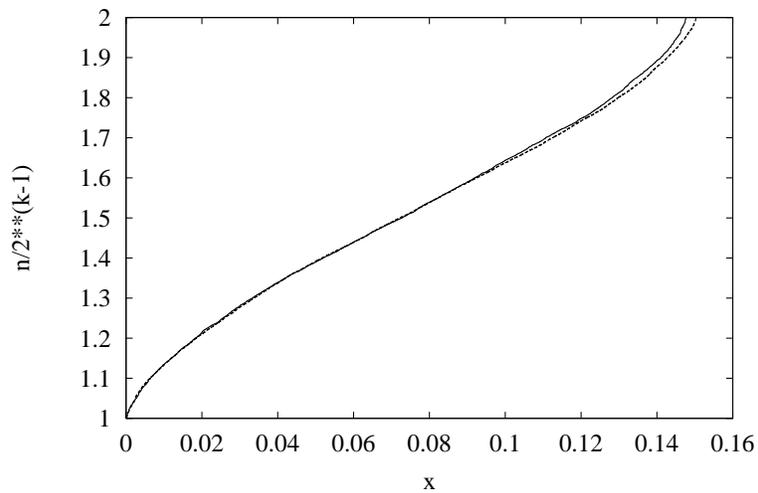}
\parbox[t]{.85\textwidth}
 {
 \caption[wieoft]
 {\label{wieoft}
\small
$i$-th left-out number in $I_k$, 
rescaled by a factor $2^{k-1}$, for $k=16$ (full line) and $k=17$ 
(dotted line).
The $x$-variable is $i$ divided by the length of interval $I_k$.
}}
\end{center}
\end{figure}

Figure \ref{wieoft} shows a plot of the $i$-th left-out number in $I_k$, 
rescaled by a factor $2^{k-1}$. 
The $x$-variable is $i$ divided by the length of interval $I_k$.
The graphs are for $k= 16$ and $17$. The difference between 
the two curves is already small. 
Of course, such graphs can also be generated for $M \neq 0$. 
They look similar. 

\section{Two Cousins of Hofstadter's Sequence}

\begin{figure}
\begin{center}
\includegraphics[height=5.5cm]{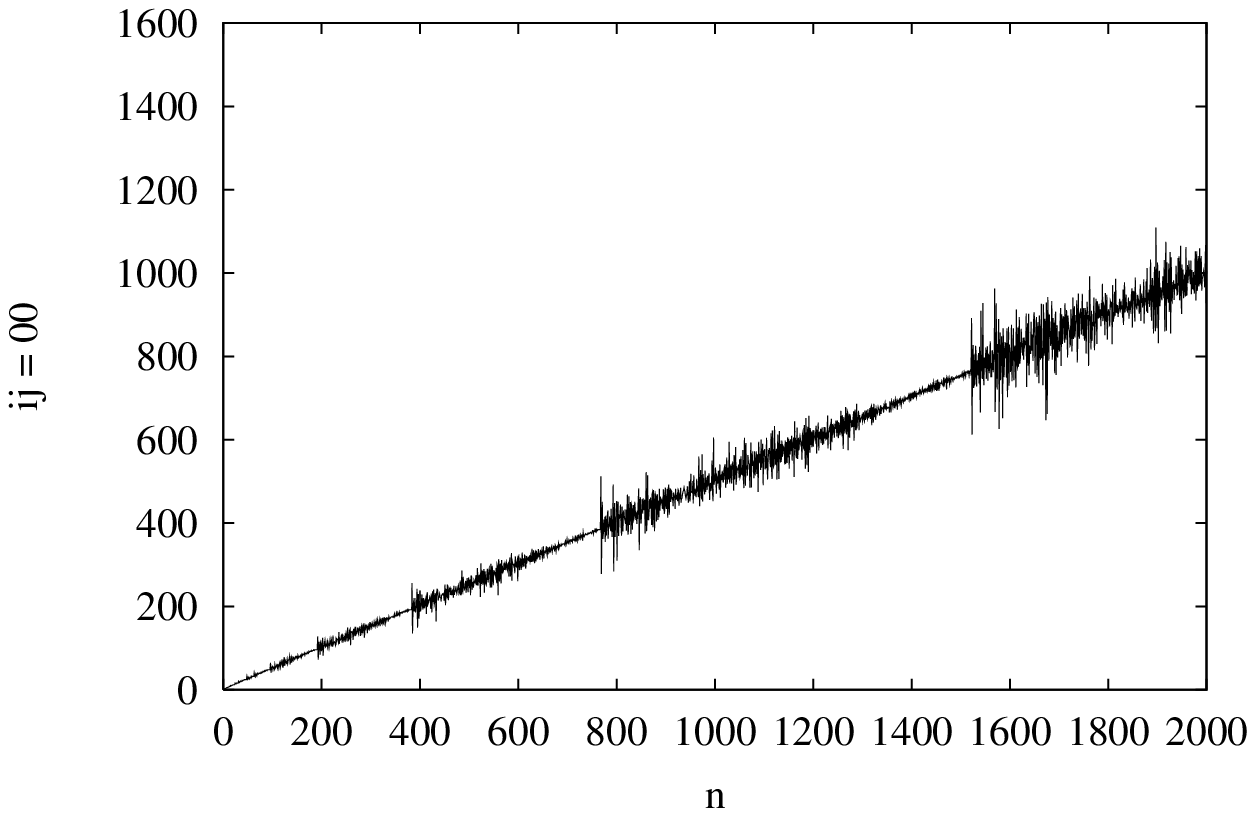}

\includegraphics[height=5.5cm]{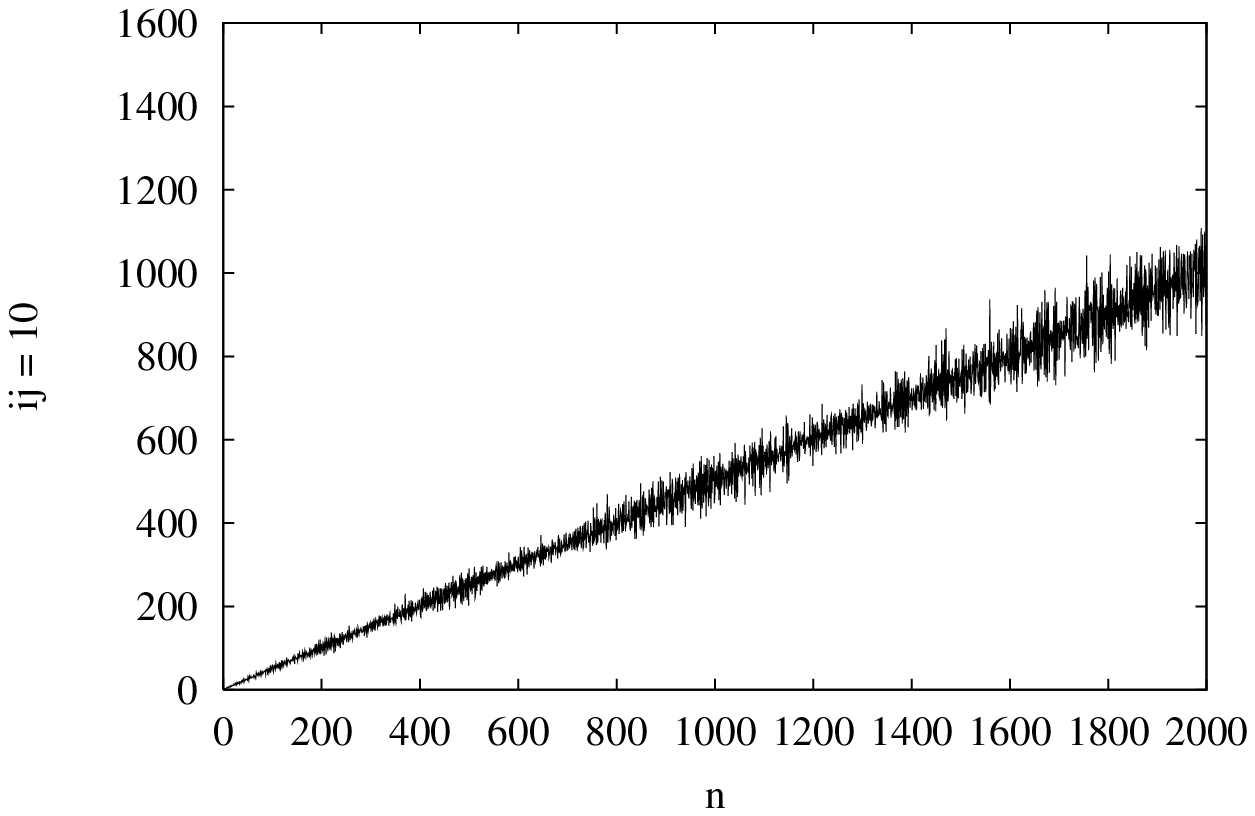}

\includegraphics[height=5.5cm]{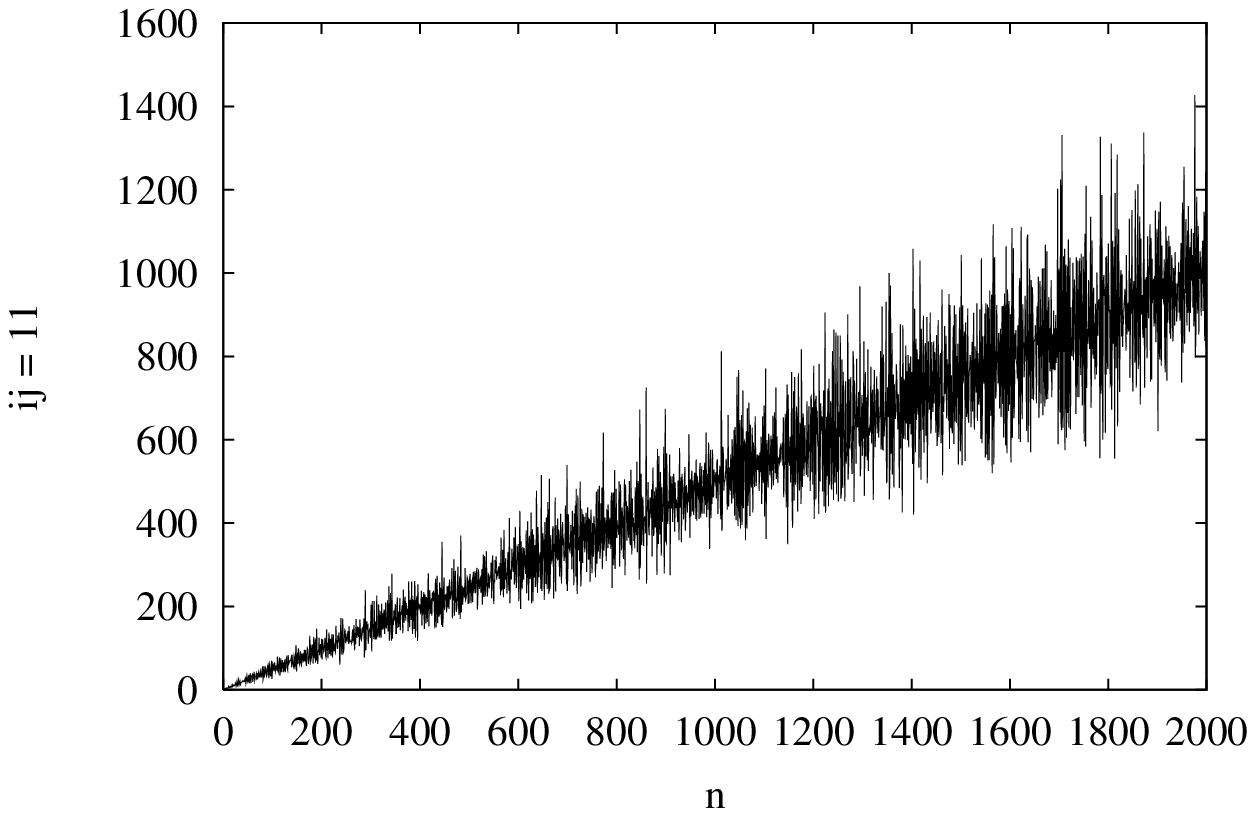}
\parbox[t]{.85\textwidth}
 {
 \caption[appearF]
 {\label{appearF}
\small
 Graphs of the sequences $F_{00}=Q$,
 $F_{10}$, and $F_{11}$.
}}
\end{center}
\end{figure}

It is natural to generalize the recurrence 
(\ref{QQQ}) by introducing constant shifts $i$ and 
$j$ in the arguments on the right hand side: 
\be\label{CCC}
\begin{array}{ll}
 & F_{ij}(n) = F_{ij}(n_1) + F_{ij}(n_2) \quad \mbox{for} ~~ n > 2 \, ,\\[3mm] 
 & F_{ij}(1) = F_{ij}(2) = 1 \, , \\[3mm] 
 & n_1 = n-i-F_{ij}(n-1) \, , \\[3mm] 
 & n_2 = n-j-F_{ij}(n-2) \, .
\end{array}
\ee 
Of course, one has to check whether the recursion  
(together with given initial conditions) leads to a well-defined 
sequence for all $n$. 
Ill-definition occurs if there exists an $n$ 
such that either $n_1$ or $n_2$ is outside of 
$[1,n-1]$. 
It turns out that definition (\ref{CCC}) is ill-defined except 
for the cases $ij=$ 00, 01, 10, and 11, where numerical 
evidence for  $n$ up to several millions makes 
consistency problems rather unlikely. 
Note that $F_{00}=Q$. The sequence with $ij=01$ seems to have 
a simple regular structure, very similar to Tanny's sequence \cite{Tanny}.
The other two cousins, $F_{10}$ and $F_{11}$, look chaotic.  A graph of
the first 2000 elements of $F_{00}$, $F_{10}$, and $F_{11}$ is shown
in figure \ref{appearF}.

\subsection{Statistical Properties}

We consider the sequences $\tilde F_{ij}(n)= F_{ij}(n)-n/2$.  
Again we study the variances 
$M(k)$, defined through
\be 
M(k)^2 =
\bigl\langle
\tilde F_{ij}(n)^2
\bigr\rangle_k
-
\bigl\langle
\tilde F_{ij}(n)
\bigr\rangle_k^2 \, ,
\ee 
where $\langle (.) \rangle_k$ denotes the average over
intervals $[2^{k-1}+1,2^k]$. 
Table~\ref{msqrF} shows the results for 
$ \alpha_k = \log_2 (M(k)/M(k-1))$, where $k \leq 25$. 
We estimate for $\alpha = \lim_{k\rightarrow \infty}= 0.88(1)$ for $ij=00$
(Hofstadter sequence), 
$0.86(1)$ for $ij=10$, and $0.89(1)$ for $ij=11$.
It seems that the exponent for $F_{10}$ is smaller than 
that for the other sequences. Note however, that 
the given error bars could be underestimated.
There are still fluctuations in table \ref{msqrF}, 
and we cannot strictly rule out the possibility 
that the exponents of the three sequences agree. 

\begin{table}
\small 
\begin{center}
\begin{tabular}{r|c|c|c}
$k$ & 00 & 10 & 11 \\ 
\hline 
 13 & 0.849 & 0.852 &  0.867 \\
 14 & 0.885 & 0.864 &  0.925 \\
 15 & 0.879 & 0.869 &  0.904 \\
 16 & 0.879 & 0.862 &  0.883 \\
 17 & 0.870 & 0.863 &  0.895 \\
 18 & 0.882 & 0.865 &  0.889 \\
 19 & 0.881 & 0.859 &  0.895 \\
 20 & 0.882 & 0.857 &  0.886 \\
 21 & 0.882 & 0.859 &  0.891 \\
 22 & 0.880 & 0.864 &  0.890 \\
 23 & 0.882 & 0.861 &  0.887 \\
 24 & 0.880 & 0.857 &  0.884 \\
 25 & 0.876 & 0.851 &  0.878 \\
\hline 
$\alpha$ & 0.88(1) & 0.86(1) & 0.89(1) \\
\hline 
 \end{tabular}
\parbox[t]{.85\textwidth}
 {
 \caption[msqrF]
 {\label{msqrF}
\small
Logarithmic variance ratios $\alpha_k$ for 
$\tilde F_{00}$, $\tilde F_{10}$, and $\tilde F_{11}$.
}}
\end{center}
\end{table}                   

\begin{figure}
\begin{center}
\includegraphics[height=7cm]{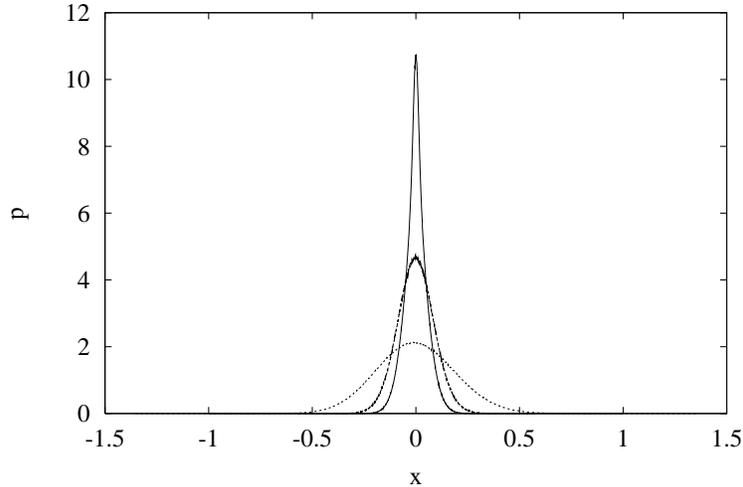}
\parbox[t]{.85\textwidth}
 {
 \caption[binF]
 {\label{binF}
\small
Statistical distributions of $\tilde F_{ij}(n)/n^\alpha$, for 
$ij=00$ (highest peak), 10, and 11 (broadest).
}}
\end{center}
\end{figure}

Figure \ref{binF} shows the statistical distribution functions 
of the quantities $\tilde F_{ij}(n)/n^\alpha$, 
where the $\alpha$'s are taken from the last line of table \ref{msqrF}. 
The binning was done over periods $[2^{k-1},2^{k}]$ for the 
10 and 11 sequences. 
For $F_{00}$ the generation structure requires intervals 
$[2^{k-1.5},2^{k-0.5}]$. The distributions for the different $k$'s 
agree nicely. 
The plot shows the $k=24$ results. The function with the highest 
peak belongs to $ij=00$, the $F_{11}$-numbers have the  broadest 
distribution.
In contrast to the 00-distribution which (as the $D$-distribution) 
goes like $\exp(-c \, x^2) /x$ for large $x$, the 
10- and 11-distributions can be fairly well approximated by 
Gaussians. It is an interesting question whether the various 
behaviours can be understood and modelled. It seems natural 
to try a fit with limiting distributions of random walks. 
Narrow non-Gaussian distribution can in principle be generated 
by sub-diffusive random walks \cite{Bouchaud}.
The observed asymptotics $\sim \exp(-c \, x^2) /x$, however, 
does not seem to be compatible with sub-diffusion.  
 
Again we observe scaling, if we look at the distributions of 
\be 
x_m = \tilde F_{ij}(n-m)- \tilde F_{ij}(n) \, . 
\ee 
More precisely, the probability density of $x_m$, $m>2$ is up 
to a rescaling the same as that of $x_{m-1}$. 
For $ij = 11$ one can detect some small scaling violations 
for the first 2 values of $m$. The approach of the $\lambda_m$ 
factors to their asymptotic value is the same for all three 
$F$-sequences, and very similar to that of the $D$-sequence. 
The convergence is again governed by a correlation length 
of 3.

\subsection{Correlation Functions}

\begin{figure}
\begin{center}
\includegraphics[height=8cm]{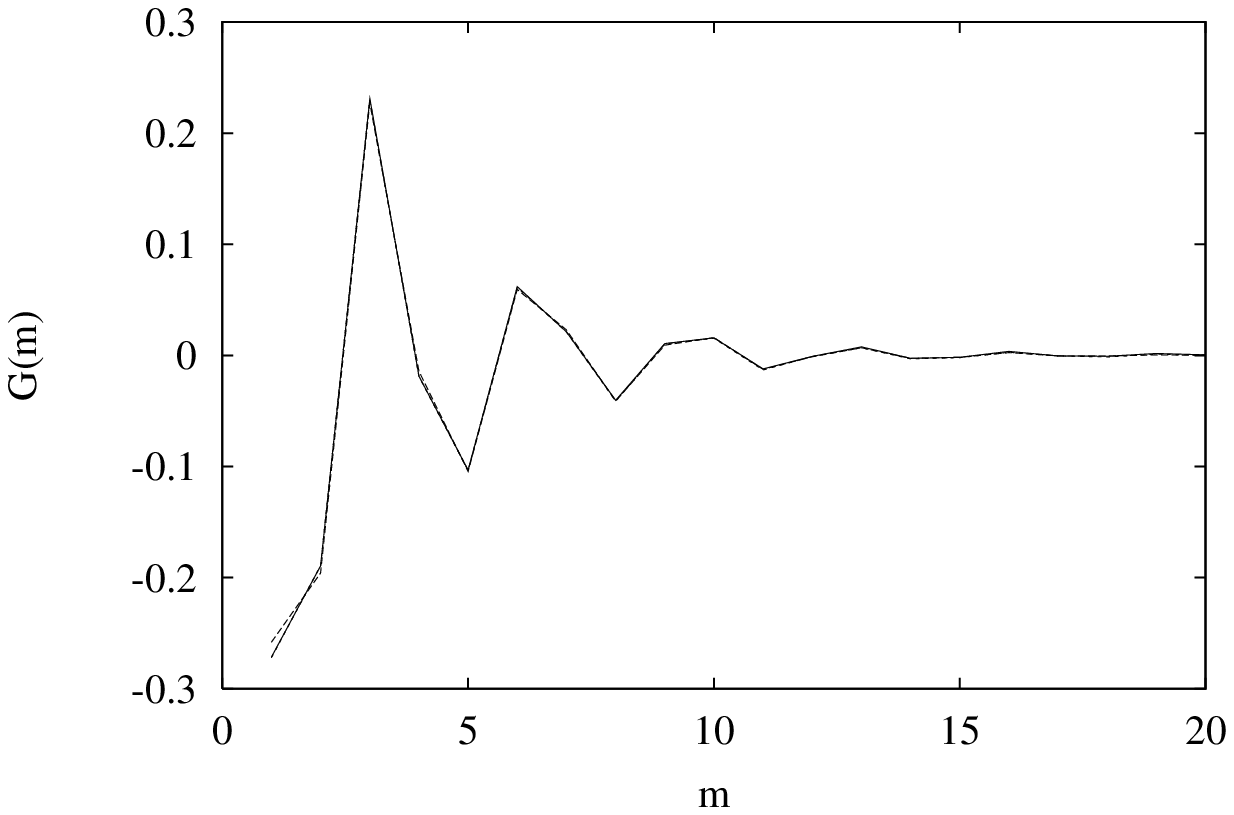}

\includegraphics[height=8cm]{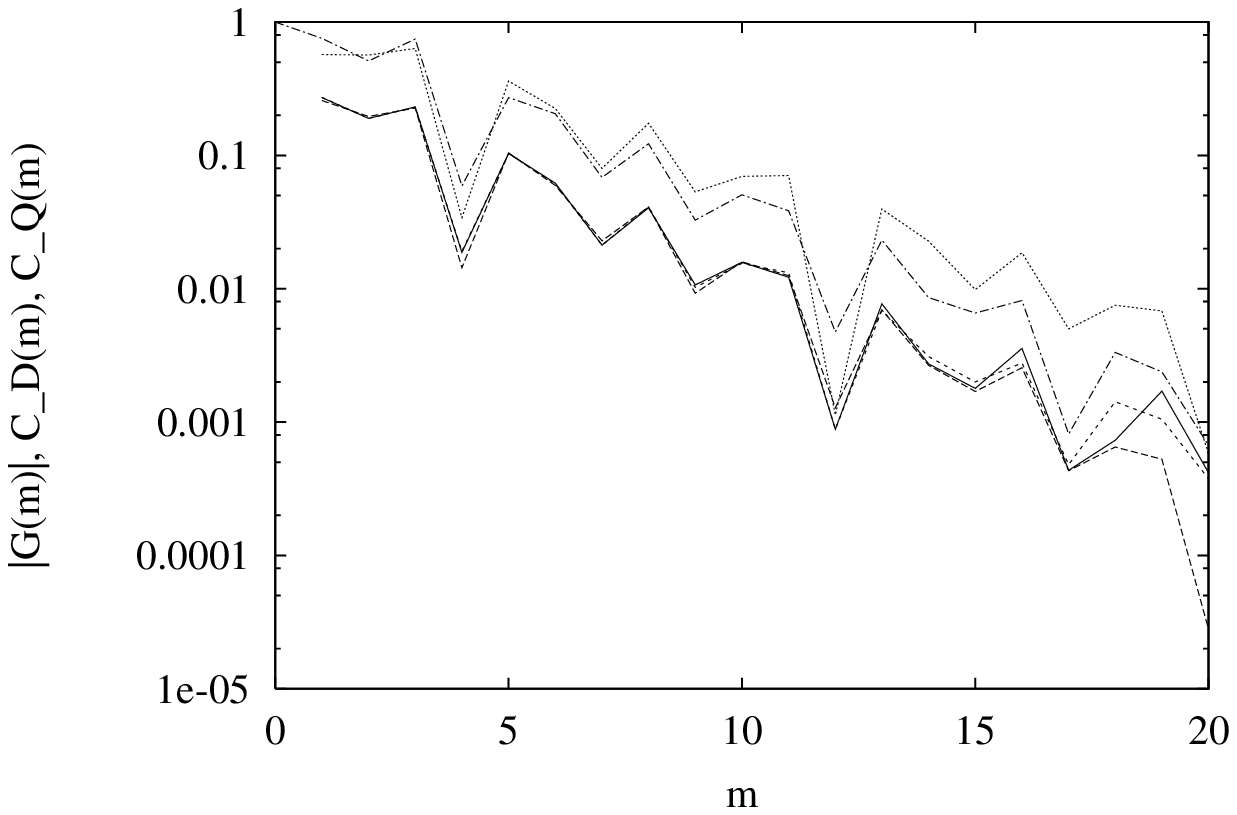}
\parbox[t]{.85\textwidth}
 {
 \caption[isicorF]
 {\label{isicorF}
\small
$G(m)$ for $ij=00$, 10, and 11 (top). 
The lower plot shows $|G(m)|$ for the same three sequences on a 
logarithmic scale (lower three graphs) 
together with the functions $C_D$ and $C_Q$ 
of figure \ref{lamdec} (upper two graphs).
}}
\end{center}
\end{figure}

For all three $F$-sequences we define a variable $\sigma_n$ through 
\be 
\sigma_n = 
\left\{ 
\begin{array}{ll}
+1 & \mbox{\ if \ } F(n) \geq n/2  \, ,  \\
-1 & \mbox{\ else} \, .
\end{array} 
\right.
\ee 
Then we ``measure'' the 2-point correlator 
\be
G(m) = \langle \sigma_n \sigma_{n-m} \rangle   
-  \langle \sigma_n \rangle^2   
\ee 
over the range $[2^{16},2^{24}]$. The results for $|G(m)|$ are shown 
in figure \ref{isicorF}. The lower part of the figure shows 
$|G(m)|$ on a logarithmic scale, together with the functions 
$C_Q$ and $C_D$ of figure \ref{lamdec}. 
The surprise is not only 
that the correlators of the three $F$-sequences seem to be identical. 
They also have a striking similarity with the functions describing
the decay of the rescaling factors $\lambda_m$.
 
\subsection{Frequency Counting}

Results for the relative frequencies of numbers $n$ occuring $M$ times
in the $F$-sequences were already given in table \ref{wieoft1t}. They
agree fairly well with those for $D$ and $Q$.

\section*{Summary and Conclusions}

In this paper, a chaotic cousin of Conway's sequence was introduced
and studied.  Its statistical properties showed some intriguing
similarities with the Hofstadter sequence $Q$ and also with the two
cousins $F_{10}$ and $F_{11}$:

\begin{itemize}
\item All the four sequences studied have (to the given precision) 
the same exponent $\alpha$, governing the increase of 
variance with increasing $n$ or $k$.
(The value for $F_{10}$ seems to 
be a little bit lower, agreement can however not be excluded.) 
\item The probability densities obey scaling. The rescaling 
parameter follows a characteristic convergence, governed by a correlation 
length 3. 
\item The correlation function $G(m)$ is identical for all three 
$F$-sequences. It also decays with correlation length 3, and in 
a way very similar to the behaviour of the $\lambda_m$ factors.
\item The relative frequencies of numbers occurring exactly $M$ times 
in the sequence seems to be the same for all the four sequences. 
\end{itemize}

In summary, the $D$-numbers and the three $F$-sequences
have a lot of common structure. One might say that they share a
universality class. A precise definition of such a class is, however,
still lacking.

The $D$-sequence is unique insofar, as it has a regular generation
structure with smooth interplays inbetween.  This could make it a
candidate for studies aiming at some rigorous results about the
chaotic recurrence relations.

It is presently an open question how much one can learn from the
relation of the $D$-recurrence with the ``solved'' $a$-sequence.  That
there is some deep relation is suggested by the apparent similarity of
the two sequences in the regions between the generations.  The
experiments with seeding the $D$-recurrence with $k$ generations of
$a$-numbers (see the end of section 2) could be a first step towards a
better understanding of this relation.

Whenever one observes the phenomenon of universality in a {\em model},
one is tempted to look for realizations of the same universality class
in {\em nature}.  It is an interesting question whether recurrences of
the type studied in this article represent real physical processes or
might be of use in the study of some dynamical system occuring in real
life. A physical picture (e.g., in terms of random walks in some
bizarre surrounding) could perhaps help to better understand some of
the interesting properties of these sequences.

\section*{Cash Prize Offered}

I offer a cash prize of \$100 to the {\em first} providing a 
rigorous proof of the claims C1, C2, and C5 stated in 
section 2.

\section*{Acknowledgements}
I would like to thank D. R. Hofstadter for interesting 
private communication.

\end{document}